\documentclass[letterpaper,conference,compsoc]{IEEEtran}
\def\version{SP}
\newif \ifsubmission \submissionfalse

\newif \iffull 
\newif \ifACM
\newif \ifUSENIX
\newif \ifIEEE
\newif \ifLNCS

\newif \ifCCS
\newif \ifSP
\newif \ifNDSS
\newif \ifCrypto
\newif \ifFC

\def\fullstring{full}
\def\ACMstring{ACM}
\def\USENIXstring{USENIX}
\def\IEEEstring{IEEE}
\def\LNCSstring{LNCS}

\def\CCSstring{CCS}
\def\SPstring{SP}
\def\NDSSstring{NDSS}
\def\Cryptostring{CRYPTO}
\def\FCstring{FC}

\ifx \version\fullstring \fulltrue \fi
\ifx \version\ACMstring \ACMtrue \fi
\ifx \version\USENIXstring \USENIXtrue \fi
\ifx \version\IEEEstring \IEEEtrue \fi
\ifx \version\LNCSstring \LNCStrue \fi

\ifx \version\CCSstring \ACMtrue \CCStrue \fi
\ifx \version\SPstring \IEEEtrue \SPtrue \fi
\ifx \version\NDSSstring \IEEEtrue \NDSStrue \fi
\ifx \version\Cryptostring \LNCStrue \Cryptotrue \fi
\ifx \version\FCstring \LNCStrue \FCtrue \fi

\ifACM \input{Conferences/ACM/acm.tex} \fi
\ifUSENIX \input{Conferences/USENIX/usenix.tex} \fi
\ifIEEE \errorcontextlines=200

\usepackage[noadjust]{cite}
\usepackage[scaled=0.88]{helvet}

 \fi
\ifLNCS \input{Conferences/LNCS/lncs.tex} \fi

\newif \ifcomments \commentsfalse
\newif \ifanon \anonfalse

\ifUSENIX \else \usepackage[table]{xcolor} \fi
\usepackage{xurl}
\usepackage{xspace}
\usepackage{hyperref}
\hypersetup{
    colorlinks=true,
    allcolors=blue,
    linkcolor=red,
    filecolor=magenta,
}
\usepackage[many]{tcolorbox}        
\usepackage{lipsum}
\usepackage{cleveref}
\usepackage{enumitem}
\usepackage{booktabs}
\usepackage{makecell}
\usepackage{multirow}
\usepackage[figuresright]{rotating}
\usepackage{array}
\usepackage{nicematrix}
\usepackage{tikz}
\usetikzlibrary{shapes, arrows.meta, positioning, calc, math, tikzmark, fit}

\usepackage{amsfonts,amsmath}
\ifLNCS\else\usepackage{amsthm}\fi
\ifACM\else\usepackage{amssymb}\fi

\newcommand{\players}{{\mathcal{P}}}
\newcommand{\player}{\ensuremath{P}}

\definecolor{ForestGreen}{RGB}{34,139,34}
\newcommand{\isgreenmoney}[1]{\textcolor{ForestGreen}{$\$#1$}}
\newcommand{\fbcontract}{\textsf{FB}\xspace}

\ifcomments
    \newcommand{\mahimna}[1]{\textsf{\small{\color{violet!80}{[Mahimna: {#1}]}}}}
    \newcommand{\kushal}[1]{\textsf{\small{\color{blue}{[Kushal: {#1}]}}}}
    \newcommand{\james}[1]{\textsf{\small{\color{green!75!black}{[James: {#1}]}}}}
    \newcommand{\ari}[1]{\textsf{\small{\color{red}{[Ari: {#1}]}}}}
    \newcommand{\jay}[1]{\textsf{\small{\color{orange}{[Jay: {#1}]}}}}
    \newcommand{\sarah}[1]{\textsf{\small{\color{red}{[Sarah: {#1}]}}}}
    \newcommand{\andres}[1]{\textsf{\small{\color{yellow!75!black}{[Andres: {#1}]}}}}
    \newcommand{\dani}[1]{\textsf{\small{\color{purple}{[Dani: {#1}]}}}}
\else
    \newcommand{\kushal}[1]{}
    \newcommand{\mahimna}[1]{}
    \newcommand{\james}[1]{}
    \newcommand{\ari}[1]{}
    \newcommand{\jay}[1]{}
    \newcommand{\sarah}[1]{}
    \newcommand{\andres}[1]{}
    \newcommand{\dani}[1]{}
\fi

\newcommand{\sysname}{Liquefaction\xspace}
\newcommand{\sk}{{\sf sk}\xspace}
\newcommand{\acma}{{\sf AM}\xspace}
\newcommand{\numrepos}{\ensuremath{m}\xspace}
\newcommand{\numtees}{\ensuremath{n}\xspace}

\newtheorem{example}{Example}

\tcbset{
    sharp corners,
    colback = white,
    before skip = 0.2cm,    
    after skip = 0.5cm      
}                           %

\newtcolorbox{boxA}{
    fontupper = \bf,
    boxrule = 1.5pt,
    colframe = black %
}

\newcommand{\mypara}[1]{\smallskip\noindent\textbf{#1}\;}

\ifLNCS\else

\newtheorem{theorem}{Theorem}

\theoremstyle{definition}

\theoremstyle{remark}
\newtheorem{remark}[theorem]{Remark}

\fi

\begin{document}

\title{Liquefaction: Privately Liquefying Blockchain Assets}

\ifSP
\IEEEoverridecommandlockouts
\makeatletter
\newcommand{\linebreakand}{%
  \end{@IEEEauthorhalign}
  \hfill\mbox{}\par
  \mbox{}\hfill\begin{@IEEEauthorhalign}
}
\makeatother
\author{
	\IEEEauthorblockN{James Austgen}
	\IEEEauthorblockA{Cornell Tech\\
		\href{mailto:james@cs.cornell.edu}{james@cs.cornell.edu}}
	\and
	\IEEEauthorblockN{Andr\'es F\'abrega}
	\IEEEauthorblockA{Cornell University\\
		\href{mailto:andresfg@cs.cornell.edu}{andresfg@cs.cornell.edu}}
	\and
	\IEEEauthorblockN{Mahimna Kelkar}
	\IEEEauthorblockA{Cornell Tech\\
		\href{mailto:mahimna@cs.cornell.edu}{mahimna@cs.cornell.edu}}
	\and
	\IEEEauthorblockN{Dani Vilardell}
	\IEEEauthorblockA{Cornell Tech\\
		\href{mailto:dv296@cornell.edu}{dv296@cornell.edu}}
	\linebreakand
	\IEEEauthorblockN{Sarah Allen}
	\IEEEauthorblockA{IC3, Flashbots\\
		\href{mailto:sarahallen@cornell.edu}{sarahallen@cornell.edu}}
	\and
	\IEEEauthorblockN{Kushal Babel}
	\IEEEauthorblockA{Cornell Tech, Monad Labs\\
		\href{mailto:babel@cs.cornell.edu}{babel@cs.cornell.edu}}
	\and
	\IEEEauthorblockN{Jay Yu}
	\IEEEauthorblockA{Stanford University\\
		\href{mailto:jyu01@stanford.edu}{jyu01@stanford.edu}}
	\and
	\IEEEauthorblockN{Ari Juels}
	\IEEEauthorblockA{Cornell Tech\\
		\href{mailto:juels@cornell.edu}{juels@cornell.edu}}
}
\else
\author{AUTHORSHIP NOT IMPLEMENTED}
\fi

\ifACM 
\begin{abstract}
Inherent in the world of cryptocurrency systems and their security models is the notion that private keys---and thus assets---are controlled by individuals or individual entities. 

We present \textit{\sysname}, a wallet platform that demonstrates the dangerous fragility of this foundational assumption by systemically breaking it. \sysname uses trusted execution environments (TEEs) to \textit{encumber} private keys, i.e., attach rich, multi-user policies to their use. In this way, it enables the cryptocurrency credentials and assets of a single end-user address to be freely rented, shared, or pooled. It accomplishes these things \textit{privately}, with no direct on-chain traces.

\sysname demonstrates the sweeping consequences of TEE-based key encumbrance for the cryptocurrency landscape. \sysname can undermine the security and economic models of many applications and resources, such as locked tokens, DAO voting, airdrops, loyalty points, soulbound tokens, and quadratic voting. It can do so with no on-chain and minimal off-chain visibility. Conversely, we also discuss beneficial applications of \sysname, such as privacy-preserving, cost-efficient DAOs and a countermeasure to dusting attacks. Importantly, we describe an existing TEE-based tool that applications can use as a countermeasure to  \sysname.

Our work prompts a wholesale rethinking of existing models and enforcement of key and asset ownership in the cryptocurrency ecosystem.

\end{abstract}

 \maketitle \pagestyle{plain}
\else \maketitle 
\begin{abstract}
Inherent in the world of cryptocurrency systems and their security models is the notion that private keys---and thus assets---are controlled by individuals or individual entities. 

We present \textit{\sysname}, a wallet platform that demonstrates the dangerous fragility of this foundational assumption by systemically breaking it. \sysname uses trusted execution environments (TEEs) to \textit{encumber} private keys, i.e., attach rich, multi-user policies to their use. In this way, it enables the cryptocurrency credentials and assets of a single end-user address to be freely rented, shared, or pooled. It accomplishes these things \textit{privately}, with no direct on-chain traces.

\sysname demonstrates the sweeping consequences of TEE-based key encumbrance for the cryptocurrency landscape. \sysname can undermine the security and economic models of many applications and resources, such as locked tokens, DAO voting, airdrops, loyalty points, soulbound tokens, and quadratic voting. It can do so with no on-chain and minimal off-chain visibility. Conversely, we also discuss beneficial applications of \sysname, such as privacy-preserving, cost-efficient DAOs and a countermeasure to dusting attacks. Importantly, we describe an existing TEE-based tool that applications can use as a countermeasure to  \sysname.

Our work prompts a wholesale rethinking of existing models and enforcement of key and asset ownership in the cryptocurrency ecosystem.

\end{abstract}

 \fi

\section{Introduction}
\label{sec:introduction}

In blockchain systems, private keys represent the ownership of assets. Each private key has full control over funds held in its associated public address. Thus, to protect against theft, participants should keep their private keys secret from others.
From this comes the widely held assumption that private keys are held by individual entities. We call this the \textit{Single-Entity Address-Ownership} (SEAO) assumption.

\mypara{The SEAO assumption.}
In the design and culture of blockchain interfaces, tools, and media, the SEAO assumption is pervasive. 
For example, address blacklists and whitelists used to categorize blockchain actors~\cite{kim2020blacklist,moser2014towards} rely on the SEAO assumption. If a single address could be used by different entities safely---i.e., without their stealing funds from one another---then the SEAO assumption would be in jeopardy. In this case, whitelists wouldn't work. An address belonging to a whitelisted individual could be used by, e.g., rented out to, an individual not on the whitelist.

Many cryptocurrency protocols and systems in fact rely critically on the SEAO assumption to achieve their security and economic models. This is true when they associate an asset, reward, or privilege non-transferably with an address, as illustrated by the following examples.

\begin{example}[Locked tokens]
\label{ex:locked}
In airdrops or token-generation events, tokens are often distributed in a \textit{locked} or \textit{unvested} state~\cite{binance_token_lockup}. Ownership is assigned to specific wallet addresses and programmed to be \textit{non-transferable} until the completion of a predefined \textit{lockup} or \textit{vesting period}.
\end{example}

\begin{example}[Loyalty rewards]
\label{ex:loyalty}
Some decentralized cryptocurrency exchanges (DEXes) offer \textit{loyalty rewards or trading discounts} to frequent traders~\cite{dydxfees:2023} or token holders~\cite{1inch:2021}. They associate rewards with particular high-activity addresses in order to incentivize use of the DEX. 
\end{example}

\begin{example}[SBTs]
\label{ex:SBTs}
\textit{Soulbound tokens} (SBTs) are special non-fungible tokens that serve as digital identity documents for individual human users~\cite{Soulbound:2022}. An SBT is intended to be ``bound to a user's soul,'' that is, assigned permanently to a particular address and held forever by a single individual. 
\end{example}

Other blockchain systems depend critically on the SEAO assumption for goals that include voting integrity---i.e., preventing vote-buying---enforcing models of token economics, ensuring transaction traceability, etc.

\subsection{\sysname}
The goal of this work is to demonstrate the inherent fragility of the SEAO assumption and its sweeping repercussions, both destructive and constructive. To this end, we introduce \textit{\sysname}, a privacy-preserving cryptocurrency wallet platform that \textit{breaks the SEAO assumption}. %

Broadly speaking, \sysname enables the transfer or \textit{liquefaction} of assets or privileges that are meant to be nontransferable / illiquid. Recalling~\Cref{ex:locked} above, a user with locked tokens can use \sysname to transfer ownership of the tokens prior to the end of their lockup or vesting period. In~\Cref{ex:loyalty} above, a user receiving loyalty discounts from a cryptocurrency exchange could use \sysname to safely rent out her address to other users, freely selling the DEX's discounts and subverting its loyalty system. Similarly, \sysname would enable users to transfer access to their SBTs in~\Cref{ex:SBTs}, completely breaking the defining ``soulbound'' property of SBTs. 

To demonstrate the extensive implications of breaking the SEAO assumption, 
\sysname enables blockchain addresses to be managed according to a rich set of access-control policies. These policies permit flexible renting, sharing, pooling, and partitioning of blockchain assets and privileges among a group of entities. They also include support for complex forms of delegation. 
\sysname operates within a blockchain-compatible trust model: Policies are enforced with high assurance by the \sysname platform, not by policy creators or other users. 

\sysname also enforces a strong notion of \textit{privacy}. \sysname produces no direct on-chain indication of which addresses it controls or what transactions it has generated. Even the creator of an address, having delegated control of resources associated with it, gains no private information about the activity of delegatees. 

In short, \sysname can \textit{privately liquefy} a wide range of assets, even if they're meant to be illiquid. \sysname demonstrates the fragility of the SEAO assumption and how this fragility can have broad repercussions for the blockchain ecosystem. Fortunately, by opening up new models of flexible, confidential asset ownership, \sysname enables not just attacks, but also beneficial new blockchain applications and markets. Critically, as we also explain, there is a practical existing countermeasure that enables blockchain applications to avoid exposure to \sysname if desired.  

\mypara{Key encumbrance.} \sysname breaks the SEAO assumption using a notion called~\textit{key encumbrance}~\cite{darkdao2018daian,kelkar2024complete}. 

An encumbered secret key \sk is one that is neither known nor managed directly by a user or administrator. Instead, \sk is generated by an application that enforces an access-control policy over the key's full lifecycle. \sysname applies such a policy to the secret keys associated with addresses it controls. (For example, if the owner of an address $A$ with secret key $\sk$ rents $A$ out, \sysname can bar the owner from accessing $\sk$ during the rental period.) 

Proposed key-encumbrance applications~\cite{matetic2018delegatee,kelkar2024complete} typically rely on the use of \textit{trusted execution environments} (TEEs) to enforce access-control policies on keys. This is our approach with \sysname.
\vspace{-2mm}

\subsection{\sysname Design and Applications}

We have implemented a demonstration version of \sysname. Its backend runs in Oasis Sapphire, a TEE-based blockchain, but \sysname can be used as a wallet for any blockchain, e.g., Ethereum. \sysname is robust to liveness failures or service discontinuation in Oasis Sapphire. 

We report on the details of our design and implementation of \sysname and explore its practical ramifications. 

\mypara{\sysname access-control policies.} A particularly important element of \sysname's design is its system of \textit{access-control policies}, which dictate how individual players can use and delegate access to an encumbered key $\sk$. To ensure broad usability, \sysname wallets work with arbitrary smart contracts, whose logic and persistent effects are computationally infeasible to reason over. It is therefore challenging to ensure that \sysname policies are both expressive---i.e., support a broad range of applications---and sound---i.e., cannot be subverted through attacks. These properties need, moreover, to be maintained as a policy changes over time to handle complex delegation histories. To address these challenges, we introduce a formal delegation model for encumbered keys that is designed to be conceptually simple, yet expressive and prevent policy conflicts.

\mypara{Adversarial ecosystem implications.} 
In breaking the SEAO assumption, \sysname erodes the security models associated with a host of applications. As another key contribution, our work is the first to enumerate these applications broadly and in many cases is the first to identify them. They include quadratic voting, soulbound tokens (SBTs), airdrops, and blockchain loyalty points. \sysname can also obfuscate transaction histories, enable stealthy wash trading, erode the fidelity of on-chain cryptocurrency analytics, and corrupt reputation-dependent systems. \sysname can also facilitate certain types of fraud, such as faking theft of funds by a known bad actor. 

For concrete study and demonstration, we focus on blockchain voting. We fully realize and report the details of a Dark DAO for voter bribery as theorized in~\cite{darkdao2018daian}, as well as a new Dark DAO ``lite'' variant realized with \sysname but allowing user participation with an ordinary (non-\sysname) wallet.

We emphasize that whether or not particular applications of \sysname constitute attacks is a subjective matter---as evidenced by, e.g., public bribery markets for blockchain voting~\cite{Eluke:2024}.

\mypara{Beneficial new applications.} Liquefying assets means creating new financial instruments, capabilities, and markets. \sysname also enables clearly beneficial applications. 

One example is a mitigation against dusting attacks, where funds are sent to an address to taint its funds. \sysname permits provable sequestering or blackholing of such funds.
Another example is privacy-preserving DAOs.
Such DAOs can circumvent detrimental leakage of intelligence---as famously exemplified by the failure of ConstitutionDAO to win a copy of the U.S. Constitution at auction due to public fundraising~\cite{Tan:2022}.

\subsection{Why Explore \sysname?}

The SEAO assumption is increasingly subject to erosion. NFT fractionalization~\cite{allen2022nfts}, liquid-staking~\cite{gogol2024sok}, and re-staking~\cite{team2024eigenlayer} have created popular new forms of liquidity for NFTs and cryptocurrency. The liquefying of governance rights in vote-buying marketplaces~\cite{lloyd2023emergent} has shown how  market forces encourage adversarial uses of liquidity. 

Some forms of liquidity can be achieved transparently using smart contracts. But as our many examples in this work show, smart contracts are of limited utility in liquefying assets, as privacy is often a critical requirement. For instance, soulbound tokens (see~\Cref{ex:SBTs}) can in principle be shared through smart contracts---but that is exactly why they are typically sent only to end-user (non-contract) addresses. 

At the same time, however, attestation-capable TEEs are pervading new computing platforms~\cite{arm_cca_2024,nvidia_confidential_computing} and blockchain systems~\cite{gogul_baliga_block_building_2023,oasis_protocol}, making private key encumbrance increasingly easy to realize. 
An inevitable outcome of this trend is that users will privately liquefy assets. \sysname sheds light on the full scope of the inevitable threats and opportunities. 

Happily, the recently proposed notion of \textit{Complete Knowledge} (CK) offers a countermeasure to key encumbrance~\cite{kelkar2024complete} where it is problematic for a blockchain application. CK enables such applications to avoid exposure to \sysname and similar tools. A key lesson of our work here is thus that as tools like \sysname increasingly realize new applications that liquefy assets, \textit{many blockchain applications will need to use CK to protect their security and economic models against SEAO-assumption breaks}. 

\subsection*{Contributions}

In showing the fragility of the SEAO assumption and the implications of breaking it, we make several contributions:

\begin{itemize}
    \item \textit{\sysname}: We introduce \sysname, a general-purpose key-encumbered wallet with rich delegation / key-encumbrance policies that shows the fragility of the SEAO assumption and the implications of breaking it. 
    \item \textit{Key-encumbrance policies}: We present the first general model for key-encumbrance policies, offering a principled approach to formulating such policies in a flexible but sound manner (\Cref{sec:liquefaction-model}). \sysname demonstrates the practicality of our model. 
    \item \textit{Implementation}: We report on a fully functional implementation of \sysname that uses Oasis Sapphire as a back end (\Cref{sec:implementation}) and ensures liveness even in the case of a service failure (\Cref{sec:liveness}). 
    \item \textit{Ecosystem impact:} We explore the potential effects of \sysname across the blockchain ecosystem (\Cref{sec:applications}). We enumerate and explore applications whose security model it can erode and report specifically on Dark DAO implementations that leverage \sysname. We also discuss use of CK as a countermeasure and, conversely, positive applications of \sysname.
\end{itemize}

We present background for our work in~\Cref{sec:background}, discuss related work in~\Cref{sec:related}, and conclude in~\Cref{sec:conclusion}.

\bigskip

\section{Background}
\label{sec:background}

\sysname uses TEE-based key encumbrance to break the SEAO assumption in cryptocurrency and decentralized finance (DeFi) systems. Here we offer brief background on these concepts.

\subsection{Cryptocurrency and DeFi}\label{subsec:defi-background}

A distinctive feature of the blockchain  ecosystem as a whole is that assets are \textit{fully controlled by means of private cryptographic (signing) keys}---in contrast to traditional web services, which offer less rigid forms of access control, e.g., through password resets. Consequently, blockchain assets are especially suitable for the type of complete programmatic control upon which \sysname is predicated.

\subsection{Trusted Execution Environments (TEEs)}
TEEs~\cite{mckeen2013innovative,mckeen2016intel} are secure areas within a computer processor that provide confidentiality and integrity for the code and data executed within them. TEEs create a protected execution environment, often implemented through extensions to the processor's instruction set architecture, which ensures that code running inside the TEE cannot be read or tampered with by unauthorized entities outside the TEE. Typical examples of TEEs include Intel SGX \cite{cryptoeprint:2016/086}, ARM Trustzone \cite{Alves2004TrustZoneI}, etc. 

TEEs have been widely integrated into blockchain systems~\cite{s23167172}, from serving as sources of randomness to supporting privacy-preserving applications.

Recent research has explored challenges associated with integrating TEEs into blockchain systems ~\cite{10.1145/3318464.3386127, DBLP:journals/corr/abs-1805-08541}. Certain attacks, such as rollback attacks or liveness failures, can compromise the security of TEE-based blockchain systems~\cite{DBLP:journals/corr/LindEKNPS17}, but certain TEE-based blockchains, such as Oasis Sapphire, include native rollback protections~\cite{jeanlouis2023sgxonerated}. %

\subsection{Key Encumbrance}

The pervasive conceptual model of key ownership today assumes that a key is controlled / owned by a single entity. For example, the private key for a cryptocurrency address is typically custodied in a user's wallet (hardware or software) and under unilateral control by the user. This model is logical in a world where sharing private keys with others exposes a user's assets or privileges to theft.

TEEs, however, can disrupt this conventional model through a notion called \textit{key encumbrance}.\footnote{Secure multi-party computation (MPC) can in principle do the same, although less practically.}  Key encumbrance eliminates direct user control of a private key \sk in favor of strictly programmatic control of \sk.
It can achieve selective delegation and fine-grained access control for private keys---and consequently blockchain assets and privileges---across a set of users. 

As an example, key encumbrance enables a briber to effectively rent a private key \sk for the purpose of voting. This is only possible if \sk is encumbered in such a way that its owner cannot vote at the same time and potentially override the briber's vote. In other words, key encumbrance in this case restricts the owner of \sk \textit{herself} from signing certain messages at certain times. \sysname implements rich policies that include restrictions of this kind.

A proposed countermeasure to key encumbrance is \textit{complete knowledge} (CK)~\cite{kelkar2024complete}---proof that a single entity / principal has direct \textit{unencumbered} knowledge of the key and consequently lacks restrictions on its use. 

\newcommand{\pk}{\mathsf{pk}}
\newcommand{\messageset}{\mathcal{M}}
\newcommand{\st}{\textsf{st}}
\newcommand{\id}{\textsf{id}}

\newcommand{\protbox}[2]{\fbox{\small\hbox{\begin{minipage}{0.97\columnwidth}\begin{center}{\bf #1}\end{center}#2\end{minipage}}}}

\section{Liquefaction Access-Control Model}
\label{sec:liquefaction-model}
In this section, we introduce a model that formalizes access policies for \sysname-encumbered keys. In our blockchain setting, since all actions require submitting appropriate \textit{signed} transactions, we will treat the ability to sign messages as a proxy for access to resources or capabilities.

\subsection{Basic Model}\label{subsec:basic-model}
We model a set $\players$ of players. Let $\messageset$ denote the space of all possible messages (e.g., a suitable subset of $\{0,1\}^*$). Intuitively, our goal is to use access-control policies to model the privilege of a particular player $\player$ in obtaining a signature (from the Liquefaction wallet) on a message $m$. For the purpose of this section, we assume the signatures are defined using a signature scheme $\Sigma = (\textsf{Gen}, \textsf{Sign}, \textsf{Ver})$.

\mypara{Access-control policy.}
An access-control policy $\Gamma$ details for each $(\player, m) \in \players \times \messageset$, whether $\player$ is allowed to obtain a signature on $m$. We allow policies to be stateful---they can depend on a general system state $\textsf{\st}$. The state $\textsf{st}$ is split into three components: (1) an internal state $\textsf{intst}$, which is kept by the Liquefaction wallet, (2) an oracle state $\textsf{ost}$ retrieved from oracle $\mathcal{O}$, which, e.g., models the current trusted blockchain state, and (3) an external state $\textsf{extst}$, which is provided by the player as part of requesting a signature. We write $\textsf{st} = (\textsf{intst}, \textsf{ost}, \textsf{extst})$.

Formally, an access-control policy is now a function $\Gamma(\player,m, \st) \to \{0,1\}$, with the semantics that $\Gamma(\player,m,\st) = 1$ denotes that $\player$ is allowed to obtain a signature on $m$, given state $\st$. We say that $\Gamma$ is \textit{stateless} if for all $\player,m, \st, \st'$, it holds that $\Gamma(\player,m,\st) = \Gamma(\player,m,\st')$. When sufficiently clear, we will simply write $\Gamma(\player,m)$ in these cases. We use $\Gamma = \textsf{0}$ and $\Gamma = \textsf{1}$ to denote the policies that approve no messages and all messages, respectively.

\mypara{Policy updates.}
We will allow for a Liquefaction wallet policy to be updated to represent changes in access. As an example, we can allow a player to transfer control of her asset to another player through the wallet itself, i.e., without transferring the asset to a different address.

To this end, we define \textit{policy update functions}---an update function $\beta$ takes as input the calling player $\player$, the current policy $\Gamma_\textsf{curr}$, the new policy $\Gamma_\textsf{new}$, and the state $\st$, and outputs a bit denoting whether the policy update (from $\Gamma_\textsf{curr}$ to $\Gamma_\textsf{new}$) is allowed. Our formalism allows for $\beta$ to be an arbitrary predicate, but we discuss some practical choices later in Section~\ref{subsec:practical-instantiations} and Section~\ref{sec:implementation}.

\begin{figure}[!t]
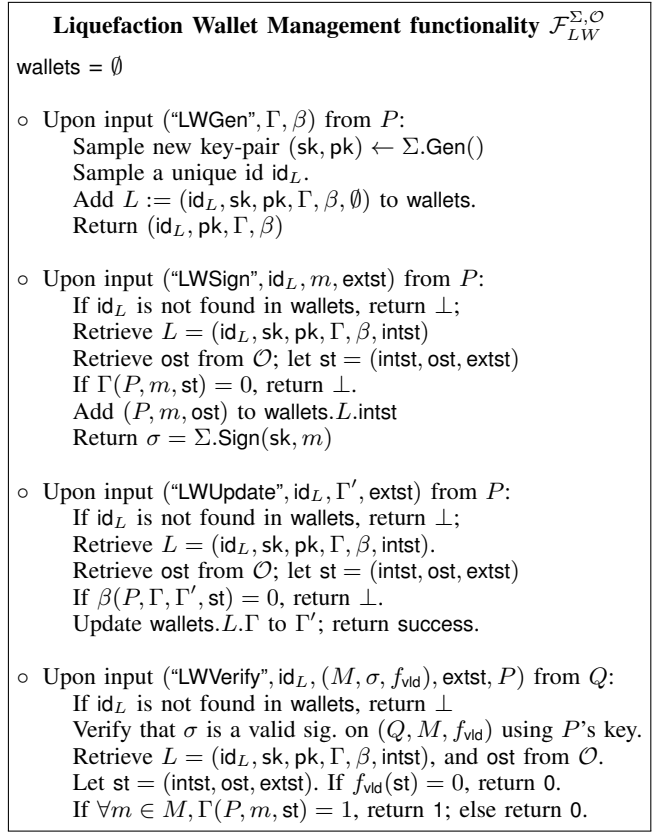

\protbox{Liquefaction Wallet Management functionality $\mathcal{F}^{\Sigma, \mathcal{O}}_{LW}$}
{    
    \textsf{wallets} = $\emptyset$ \\ 

    $\circ\:$ Upon input $(\textsf{``LWGen''}, \Gamma, \beta)$ from $P$: \\
    \hspace*{2em} Sample new key-pair $(\sk, \pk) \gets \Sigma.\textsf{Gen}()$ \\
    \hspace*{2em} Sample a unique id $\id_L$. \\
    \hspace*{2em} Add $L := (\id_L, \sk, \pk, \Gamma, \beta, \emptyset)$ to $\textsf{wallets}$.\\
    \hspace*{2em} Return $(\id_L, \pk, \Gamma, \beta)$ \\

    $\circ\:$ Upon input $(\textsf{``LWSign''}, \id_L, m, \textsf{extst})$ from $P$: \\ 
    \hspace*{2em} If $\id_L$ is not found in $\textsf{wallets}$, return $\bot$; \\ 
    \hspace*{2em} Retrieve $L = (\id_L, \sk, \pk, \Gamma, \beta, \textsf{intst})$\\
    \hspace*{2em} Retrieve \textsf{ost} from $\mathcal{O}$; let $\st = (\textsf{intst}, \textsf{ost}, \textsf{extst})$ \\
    \hspace*{2em} If $\Gamma(P, m, \st) = 0$, return $\bot$. \\
    \hspace*{2em} Add $(P,m, \textsf{ost})$ to $\textsf{wallets}.L.\textsf{intst}$\\
    \hspace*{2em} Return $\sigma = \Sigma.\textsf{Sign}(\sk, m)$ \\

    $\circ\:$ Upon input $(\textsf{``LWUpdate''}, \id_L, \Gamma', \textsf{extst})$ from $P$: \\ 
    \hspace*{2em} If $\id_L$ is not found in $\textsf{wallets}$, return $\bot$; \\ 
    \hspace*{2em} Retrieve $L = (\id_L, \sk, \pk, \Gamma, \beta, \textsf{intst})$. \\
    \hspace*{2em} Retrieve \textsf{ost} from $\mathcal{O}$; let $\st = (\textsf{intst}, \textsf{ost}, \textsf{extst})$ \\
    \hspace*{2em} If $\beta(P, \Gamma, \Gamma', \st) = 0$, return $\bot$. \\
    \hspace*{2em} Update $\textsf{wallets}.L.\Gamma$ to $\Gamma'$; return $\textsf{success}$. \\

    $\circ\:$ Upon input $(\textsf{``LWVerify''}, \id_L, (M, \sigma, f_\textsf{vld}), \textsf{extst}, P)$ from $Q$: \\
    \hspace*{2em} If $\id_L$ is not found in $\textsf{wallets}$, return $\bot$ \\
    \hspace*{2em} Verify that $\sigma$ is a valid sig. on $(Q, M, f_\textsf{vld})$ using $P$'s key. \\
    \hspace*{2em} Retrieve $L = (\id_L, \sk, \pk, \Gamma, \beta, \textsf{intst})$, and \textsf{ost} from $\mathcal{O}$.  \\
    \hspace*{2em} Let $\st = (\textsf{intst}, \textsf{ost}, \textsf{extst})$. If $f_\textsf{vld}(\st) = 0$, return $\textsf{0}$. \\
    \hspace*{2em} If $\forall m \in M, \Gamma(P, m, \st) = 1$, return \textsf{1}; else return $\textsf{0}$.
}
\caption{Liquefaction Wallet Management Functionality}
\label{fig:lwmanager}
\end{figure}

\mypara{Liquefaction wallet.}
Intuitively, a Liquefaction wallet is a key-pair $(\sk, \pk)$ associated with a particular policy $\Gamma$, and an update function $\beta$. In~\Cref{fig:lwmanager}, we specify a general wallet manager functionality that can create, interact with, and update wallets. When an operation is performed, it is implicitly assumed that the functionality authenticates the interaction with the player. The manager functionality includes the following operations:
\begin{itemize}
    \item $\textsf{LWGen}$ creates a new Liquefaction wallet with a fresh encumbered key-pair $(\sk, \pk)$, where $\sk$ is only held by the manager functionality, and thus encumbered. Access-control management is specified through $\Gamma$ and $\beta$ as discussed previously.
    
    \item $\textsf{LWSign}$ allows a player with the correct access to obtain a signature on a message $m$ from the wallet's encumbered key.

    \item $\textsf{LWUpdate}$ allows for updates to the access policy $\Gamma$, as specified by $\beta$.

    \item $\textsf{LWVerify}$ allows a player chosen by $\player$ to verify that $\player$ has some specified access (given by message set $M$) to the wallet in particular states---given by a validity function $f_\textsf{vld}(\st) \to \{0,1\}$.
\end{itemize}

\begin{remark}[Liquefying Liquefaction]
Note that a player $\player$'s access within a \sysname wallet could itself be ``liquefied'' by $\player$---i.e., controlled through a separate \sysname instance and partitioned among a different set of players without the knowledge of the wallet itself. As an example, given access for a message set $M$, $\player$ could partition this further on her own by giving $Q$ access to $M'$ and $R$ access to $M \setminus M'$. Intuitively, this will be done by $\player$ encumbering the key she uses to authenticate to the Liquefaction wallet. 

Such hierarchical liquefaction can be prevented if the \sysname wallet uses proofs of complete knowledge for authentication (see Section~\ref{sec:complete-knowledge} for details).

\end{remark}

\subsection{Practical Instantiations of \texorpdfstring{$\Gamma$}{Gamma} and \texorpdfstring{$\beta$}{Beta}}
\label{subsec:practical-instantiations}

While our formalism allows for significant generality in \sysname wallets, important practical considerations arise in practice. We describe some of these in this section, and detail how they influence our choices for permissible policies $\Gamma$ and for the policy-update function $\beta$.

First, a tension exists between the expressiveness of policies in \sysname and the feasibility for \sysname to identify or rule out potential policy conflicts. For example, the policy-update function $\beta$ might permit an extremely expressive range of policies---e.g., policies realizable as arbitrary code. In this case, though, then determining whether a policy update introduces conflicts in asset control---i.e., conflicting ownership claims over assets---would be uncomputable in general. Consequently, $\beta$ must bound the expressiveness of permissible policies.

A second important practical consideration is \textit{conceptual simplicity}. The wallet policy and how it can change must be easy for ordinary users to understand. Only then can users reason about and benefit from applications built on \sysname. For example, a user lending an asset in a \sysname wallet has the simple assurance that she will regain control of the asset unconditionally after a predetermined interval of time elapses (as opposed to depending on complex  conditions).  

As a reasonable middle-ground, our implementation relies on a policy principle called \textit{asset-time segmentation}.

\mypara{Key principle for policies $\Gamma$: Asset-time segmentation.} As a blockchain wallet system, \sysname manages control over blockchain assets (e.g., cryptocurrency or tokens). 
\sysname follows a key guiding principle for any policy: \textit{a given asset is controlled exclusively by a single player $\player$ at any given time}. This basic principle is what we refer to as asset-time segmentation.

While asset-time segmentation is conceptually simple, 
one subtlety arises in the case of fungible assets, e.g., ether in Ethereum. To avoid unduly inflexible policies that must allocate an entire asset type to a single player, \sysname permits individual holdings to be shared among multiple players as distinct assets. These individual holdings may be thought of as virtual sub-accounts. 

For instance, a policy for a wallet holding 15 ETH may specify that Alice controls the asset 5 ETH over the next week, meaning that she may spend up to 5 ETH during that time. Bob may simultaneously control 10 ETH during a concurrent interval of time. We regard their two allocations as distinct assets, realizable as a partitioning of the ETH balance in the wallet. 

Formally, suppose that that we have a set $\textbf{A} = \{\mathsf{A}_1, \dots, \mathsf{A}_k\}$ of assets. Let $\textsf{spent}(\player, \textsf{intst}, \mathsf{A}_i) \to \mathbb{Z}^{\geq 0}$ denote the quantity of asset $\mathsf{A}_i$ currently spent by $\player$. For a message $m$, let $\textsf{spend-req}(m, \mathsf{A}_i) \to \mathbb{Z}^{\geq 0}$ denote that quantity of asset $\mathsf{A}_i$ that $m$ is requesting to spend. 

Now, we can define an asset-time segmented policy $\Gamma$ as below:
\begin{itemize}
    \item $\Gamma$ is associated with a balance function $\textsf{bal} : \players \times \textbf{A} \to \mathbb{Z}^{\geq 0}$ and a time function $T : \players \times \textbf{A} \to \mathbb{Z}^+ \cup \infty$.
    \item A player $\player$ has access only to her allocated asset balance; that is, $\Gamma(\player, m, \st) = 1$ iff for all $\mathsf{A}_i$, it holds that $\textsf{spent}(P, \textsf{intst}, \mathsf{A}_i) + \textsf{spend-req}(m , \mathsf{A}_i) \leq \textsf{bal}(\player, \mathsf{A}_i)$.
    \item Access is provided only until a specified expiration time. That is, whenever $\textsf{spend-req}(m, \mathsf{A}_i) > 0$ and $t > T(P, \mathsf{A}_i)$, it holds that $\Gamma(\player, m, \st) = 0$. Here, $t$ denotes the current time (provided by $\mathcal{O}$).
\end{itemize}

\mypara{Policy-update function $\beta$.}
Intuitively, a player $\player$ should only be able to update policies for assets that she currently controls; she should not be able to arbitrarily modify access currently held by other players. 

Formally, a policy update $\Gamma \to \Gamma'$ will be allowed, or in other words, $\beta(P, \Gamma, \Gamma', \st) = 1$, only if the following hold:
\begin{itemize}
    \item $\Gamma'$ is also an asset-time segmented policy.
    
    \item $\player$ cannot revoke access held by other players. That is, $[\Gamma(Q, m, \st) = 1] \Rightarrow [\Gamma'(Q, m , \st) = 1]$ for $Q \neq P$.
    
    \item Any new access given to a different player $Q$ comes from $\player$. That is, if $\Gamma(Q, m, \st) = 0$ and $\Gamma'(Q, m, \st) = 1$, then it must be that $\Gamma(P, m, \st) = 1$.
\end{itemize}
Further restrictions may be added to $\beta$ as required by the specific application.

\mypara{Pre-signing considerations.}
Suppose that a \sysname wallet updates its policy from $\Gamma$ to $\Gamma'$ such that a player $P$ has access to asset $\mathsf{A}$ in $\Gamma$, but not in $\Gamma'$. 
Care must be taken then to ensure that $\player$ did not obtain a signature $\sigma$ on a message $m$ while $\Gamma$ is active, i.e., \textit{before} the update to $\Gamma'$, such that $\sigma$ corresponds to a transaction enabling access to $\mathsf{A}$ even \textit{after} the update. This will be checked within $\beta$. 

As an example, if not correctly handled, pre-signing could allow $P$ to double-spend as follows. $P$ obtains (but does not use) a signature $\sigma$ from the \sysname wallet to transfer an asset $\mathsf{A}$ to a different address it controls. $P$ then sells its access to $\mathsf{A}$ to another player $Q$ and updates $\Gamma \to \Gamma'$ to internally switch the owner of $\mathsf{A}$ to $Q$. Finally, $P$ submits $\sigma$ to the blockchain to transfer $\mathsf{A}$ out of the wallet---stealing it out from under $Q$.

While application-specific semantics dictate how we need to handle pre-signing concerns, we give a general design intuition for $\beta$ here. 
Ignoring balances for simplicity, for each player $\player$ and asset $\mathsf{A}_i$, $\beta$ will determine, based on the state $\st$, whether the asset needs to be ``sealed''---that is, either it has already been spent, or $\player$ has been provided a signature on a message that \emph{could} spend it. Sealed assets may be unsealed later according to application semantics. Now, an update $\Gamma \to \Gamma'$ will only be allowed if it changes access control solely for ``unsealed'' assets. Assets with balances can be handled by  determining what portion of the balance is unsealed. 

Section~\ref{subsec:transaction-encumbrance-policy} provides more details specific to our implementation.

\subsection{Privacy}

\sysname aims not just to enforce access-control policies correctly, but to do so in a
privacy-preserving way. Specifically, a player $\player$ should learn only her own portion of a given wallet policy $\Gamma$, i.e., only $\Gamma(P,m, \textsf{st})$, and not $\Gamma(Q, m, \textsf{st})$ for any other player $Q$. The only way for $P$ to learn any information about $\Gamma(Q, m, \textsf{st})$ should be through $\textsf{LWVerify}$---and only upon authorization by $Q$.  

Formally, we define \textit{policy privacy} as a player $P$ not being able to distinguish whether the current policy is $\Gamma$ or $\Gamma'$ when they give identical access to $\player$, i.e., when $\Gamma(P,m,\st) = \Gamma'(P,m,\st)$ for all $(m,\st)$.

Note that since the assets held by a \sysname wallet are still publicly visible on-chain, policy privacy cannot prevent anyone from observing on-chain transactions. Rather, it conceals players' control of assets within the wallet.

\section{Implementation}
\label{sec:implementation}

To illustrate that Liquefaction is practical, we have implemented a wallet prototype which conforms to the core functionality described in~\Cref{sec:liquefaction-model}. Our wallet prototype is a smart contract written in the most popular smart-contract programming language, Solidity. Despite its reliance on TEEs, our prototype demonstrates that developing \sysname tools and key-encumbrance policies need not require special knowledge of TEEs, thanks to the abstractions provided by TEE-based blockchains. In this section, we describe our implementation and its properties. Our open-source implementation can be found at\begingroup
\def\UrlBreaks{\do\/\do-}
\url{https://github.com/key-encumbrance/liquefaction}.
\endgroup

Our wallet prototype operates on Oasis Sapphire, a blockchain implementing the Ethereum Virtual Machine which runs entirely inside a TEE.
As a result, Sapphire supports smart contracts whose state is kept private both during execution (in memory) and after execution (in storage).

A function can be called on a Sapphire smart contract in one of two ways: (1) by submitting a blockchain transaction, or (2) by making a free, off-chain query that simulates a blockchain transaction without modifying persistent storage. Both methods require a signed message for authentication and run entirely inside the TEE of an Oasis Sapphire node.

\subsection{Wallet Architecture}
\label{subsec:wallet-architecture}
An instance of \sysname, which we call a \textit{wallet}, is realized as a contract on the TEE blockchain that we call the \textit{wallet contract}.
At its core, a wallet relies on several components: the backing TEE-based blockchain which provides built-in authentication, timestamping, and the platform on which our wallet contract runs; the wallet contract itself, which stores key material and directly signs messages; specialized, time-bound contracts that enforce policies assigning control of assets; and an \textit{access manager}, authorized to initiate policy transitions.

\mypara{Access manager.} Use of the wallet begins with a player (or smart contract) requesting that the wallet contract generate a fresh encumbered key $\sk$ using the TEE blockchain's built-in entropy generation methods. The player or smart contract who initiates this request is designated the \textit{access manager} ($\acma$) of $\sk$, serving as the authority for initiating policy transitions. For example, Alice might set up an encumbered account to sell her voting power in a DAO. She would then serve as the access manager for $\sk$. She can then decide, for example, whether she wants to accept a particular bribe and assign her voting rights to the counterparty. We emphasize, however, that the encumbered key $\sk$ is kept secret at all times within the wallet contract. 

The $\acma$ can alternatively itself be a smart contract. In this case, $\sk$ is exclusively under programmatic control. We refer to the $\acma$ in this case as an \textit{autonomous wallet}. (It may be thought of loosely as a private DAO. In fact one of our Dark DAO implementations is realized using an autonomous wallet.)

\mypara{Access control via sub-policies.} As explained in Section~\ref{sec:liquefaction-model}, use of $\sk$ is governed by an access control policy. In our prototype, access control policies are realized via a collection of \emph{sub-policies}. Each sub-policy is implemented as a separate smart contract called a \textit{policy contract}. A sub-policy enforces control of a given asset by a user or users. That is, it authorizes signatures on the asset when called by authorized users, as we explain below. Sub-policies expire after a pre-determined amount of time, which determines the duration of the access granted by the sub-policy. Together, sub-policies encode the complete access control policy associated with $\sk$.

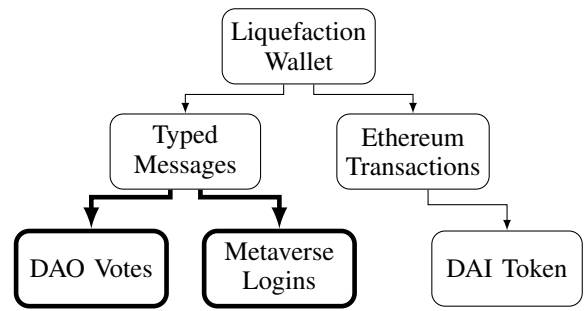
\begin{figure}[b]
	\centering
	\scalebox{1}{
		\tikzset{
    policyBox/.style = {
        rounded corners=5pt,
        minimum width=2cm,
        minimum height=1cm,
        draw
    },
    policyPoofed/.style = {
        policyBox, %
        fill=none, %
        dashed, %
        dash pattern=on 1pt off 3pt, %
        color=black!40 %
    },
    veryBold/.style = {
        line width=1.75pt
    },
}

\begin{tikzpicture}[]
    \node[policyBox, align=center] (root) at (0,0) {Liquefaction \\ Wallet};
    \coordinate[below=5pt of root, left=0.2] (rootOffset1);
    \coordinate[below=5pt of root, right=0.2] (rootOffset2);
    
    \node[policyBox, align=center, anchor=east] (typedMessages) at (-0.5,-1.4) {Typed \\ Messages};
    \node[policyBox, align=center, anchor=west] (ethTransactions) at (0.5,-1.4) {Ethereum \\ Transactions};
    
    \node[policyBox, veryBold, align=center] (daoVotes) at ([yshift=-1.05cm, xshift=-1.25cm]typedMessages.south) {DAO Votes};
    \node[policyBox, veryBold, align=center] (metaverseLogin) at ([yshift=-1.05cm, xshift=1.25cm]typedMessages.south) {Metaverse \\ Logins};
    \node[policyBox, align=center] (daiToken) at ([yshift=-1.05cm, xshift=1.25cm]ethTransactions.south) {DAI Token};
    
    \draw[->, >=latex] (rootOffset1) -- ++(0,-0.133) -| (typedMessages.north);
    \draw[->, >=latex] (rootOffset2) -- ++(0,-0.133) -| (ethTransactions.north);
    
    \draw[veryBold, ->, >=latex] ([xshift=-0.2cm]typedMessages.south) -- ++(0,-0.1) -| (daoVotes.north);
    \draw[veryBold, ->, >=latex] ([xshift=0.2cm]typedMessages.south) -- ++(0,-0.1) -| (metaverseLogin.north);
    \draw[->, >=latex] ([xshift=0.2cm]ethTransactions.south) -- ++(0,-0.2) -- ++(1.05,0) -- (daiToken.north);
\end{tikzpicture}
	}
	\caption{Example showing how new sub-policies are spawned. The Typed Messages sub-policy adds the DAO Votes and Metaverse Login sub-policies to the wallet's delegation tree---shown in bold for emphasis.}
    \label{fig:subpolicy-add}
\end{figure}

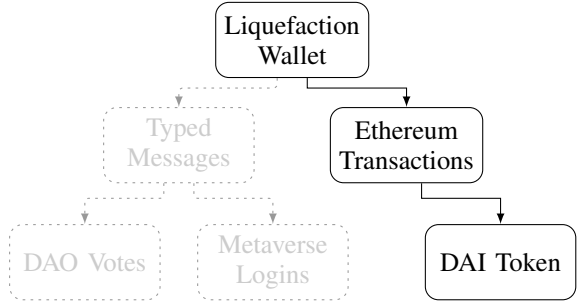
\begin{figure}
	\centering
	\scalebox{1}{
		\tikzset{
    policyBox/.style = {
        rounded corners=5pt,
        minimum width=2cm,
        minimum height=1cm,
        draw
    },
    policyPoofed/.style = {
        policyBox, %
        fill=none, %
        dashed, %
        dash pattern=on 1pt off 2pt, %
        color=black!40, %
        text=black!20,
        line width=0.4pt
    },
    poofedArrow/.style = {
        policyPoofed,
        rounded corners=1.5pt
    },
    veryBold/.style = {
        line width=1.75pt
    },
}

\begin{tikzpicture}[]
    \node[policyBox, align=center] (root) at (0,0) {Liquefaction \\ Wallet};
    \coordinate[below=5pt of root, left=0.2] (rootOffset1);
    \coordinate[below=5pt of root, right=0.2] (rootOffset2);
    
    \node[policyPoofed, align=center, anchor=east] (typedMessages) at (-0.5,-1.4) {Typed \\ Messages};
    \node[policyBox, align=center, anchor=west] (ethTransactions) at (0.5,-1.4) {Ethereum \\ Transactions};
    
    \node[policyPoofed, align=center] (daoVotes) at ([yshift=-1.05cm, xshift=-1.25cm]typedMessages.south) {DAO Votes};
    \node[policyPoofed, align=center] (metaverseLogin) at ([yshift=-1.05cm, xshift=1.25cm]typedMessages.south) {Metaverse \\ Logins};
    \node[policyBox, align=center] (daiToken) at ([yshift=-1.05cm, xshift=1.25cm]ethTransactions.south) {DAI Token};
    
    \draw[poofedArrow, ->, >=latex] (rootOffset1) -- ++(0,-0.133) -| (typedMessages.north);
    \draw[->, >=latex] (rootOffset2) -- ++(0,-0.133) -| (ethTransactions.north);
    
    \draw[poofedArrow, ->, >=latex] ([xshift=-0.2cm]typedMessages.south) -- ++(0,-0.2) -| (daoVotes.north);
    \draw[poofedArrow, ->, >=latex] ([xshift=0.2cm]typedMessages.south) -- ++(0,-0.2) -| (metaverseLogin.north);
    \draw[->, >=latex] ([xshift=0.2cm]ethTransactions.south) -- ++(0,-0.2) -- ++(1.05,0) -- (daiToken.north);
\end{tikzpicture}
	}
	\caption{Example showing how sub-policies lose access to signatures no later than their parents. The Typed Messages policy expires; consequently, its sub-policies also lose access.}
    \label{fig:subpolicy-expiration}
\end{figure}

When an encumbered key is first generated, no signing privileges are available by default (thus, $\Gamma = \textsf{0}$). To grant access to the key, $\acma$ authorizes a sub-policy to sign over a set of messages. Further modifications to signing capabilities require the creation of new sub-policies, which must not conflict with existing ones. This implies that policy updates can happen in only two ways: updates to access control arise from the \textit{addition} of new sub-policies that are compatible with existing ones (\Cref{fig:subpolicy-add}) or from the \textit{expiration} of existing sub-policies (\Cref{fig:subpolicy-expiration}). Access to an asset, once granted to a party, cannot be revoked while the corresponding sub-policy is active.

There is only one way for sub-policies to be added to a wallet: Sub-policies may spawn further, dependent sub-policies, representing more fine-grained delegation of assets. Sub-policies thus form a tree, arranged hierarchically, such that the leaves are non-conflicting. 

\mypara{Policy transitions.}
A wallet policy $\Gamma$ encompasses a set of sub-policies; thus $\Gamma = \{\Gamma_1, \Gamma_2, \Gamma_3, \ldots\, \Gamma_n\}$ for some $n$, where each sub-policy $\Gamma_j$ corresponds to a node in the sub-policy delegation tree for the wallet, with $\Gamma_1$ at the root. As sub-policies change through creation and expiration, the result is that a wallet transitions among a sequence of different policies $\Gamma^{(1)}, \Gamma^{(2)}, \Gamma^{(3)}, \ldots$, each with its own distinct set of sub-policies, i.e., $\Gamma^{(i)} = \{\Gamma^{(i)}_1, \Gamma^{(i)}_2, \ldots, \Gamma^{(i)}_{n_i}\}$. 

The policy-update function $\beta$ determines the validity of a policy transition $\Gamma^{(i)} \rightarrow \Gamma^{(i+1)}$ when new sub-policies are created. Each sub-policy has its own local update function to decide which sub-policies may be created through delegation as children in the tree. As $\beta$ only permits the instantiation of valid new sub-policies (with compliant policy contracts), \sysname thus enforces $\beta$ within sub-contracts themselves, i.e., a sub-policy enforces validity under $\beta$ of the new sub-policies it spawns.

The key property enforced globally by $\beta$ in policy transitions is \textit{asset-time segmentation}, as discussed in~\Cref{subsec:practical-instantiations}. This principle is enforced by assigning a given sub-policy \textit{exclusive} access to a given asset until that access lapses at the policy’s expiration time. Thus, a \sysname wallet (including $\textsf{AM}$) cannot remove or reduce access already conferred: Once Alice agrees to hand off her voting rights, she cannot also, for instance, assign them to herself before the agreement expires. Asset-time segmentation also prevents conflicts between sub-policies---e.g., Alice selling a single voting capability to two different bribers. \Cref{fig:time-asset-seg} illustrates how this works.

\begin{figure}
    \centering
    \includegraphics[width=\linewidth]{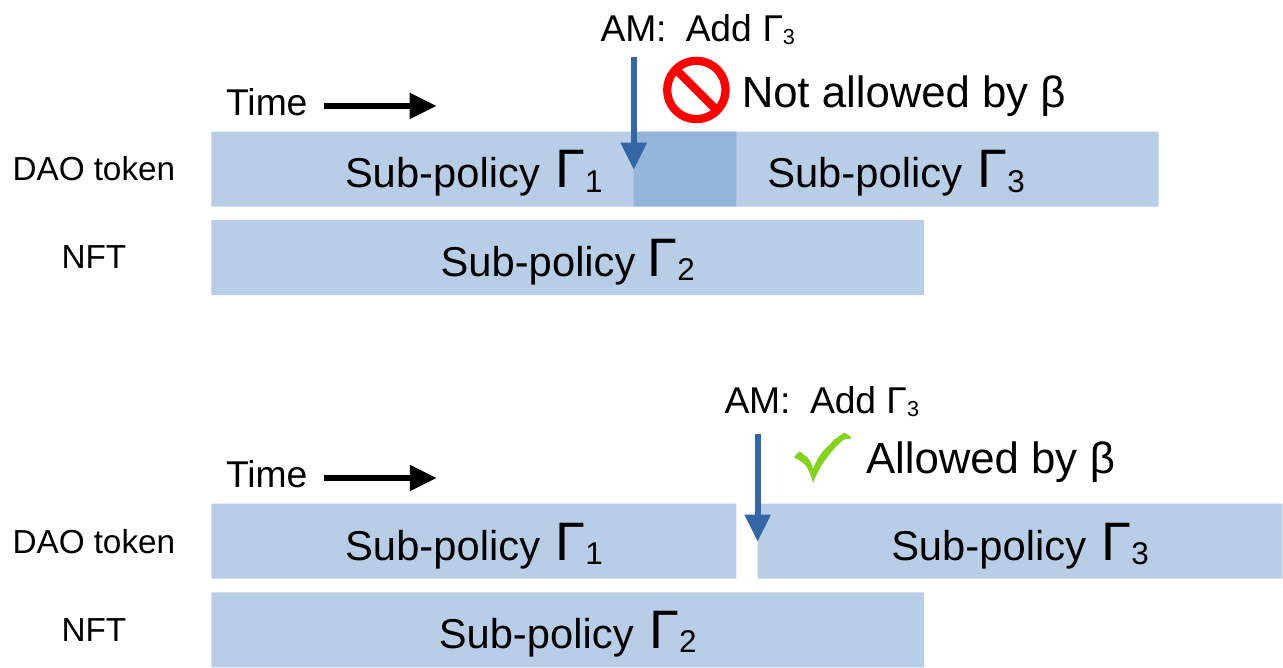}
    \caption{Asset-time segmentation requires assets be issued exclusively to a single sub-policy at a time.} 
    \label{fig:time-asset-seg}
\end{figure}

\mypara{Transaction signing.} A blockchain transaction takes the form of a signature with private key $\sk$ on a transaction message $m$. A sub-policy $\Gamma_j$ authorizes specific users to sign transaction messages for particular assets.

To authorize a transaction, a user $\player$ sends a message $m$ that she wants to have signed to the policy contract for the pertinent sub-policy $\Gamma_j$. The message-signing request is then checked recursively: first, the policy contract for $\Gamma_j$ verifies that $\Gamma_j(\player, m, \textsf{st}) \stackrel{?}{=} 1$; then the parent sub-policy in the delegation tree checks compliance, etc. Upon successful verification by the root sub-policy $\Gamma_1$, the wallet contract generates a signature on message $m$ and passes it down the delegation tree back to the policy contract for $\Gamma_j$. 

\Cref{fig:liq-wallet-contract-flow} illustrates this process for an example sequence of sub-policies forming a path in a wallet's delegation tree. The specific sub-policies there are immaterial to the signing process illustrated in the example, but look ahead to the Dark DAO application discussed in~\Cref{subsec:governance}.

\begin{figure}
	\centering
	\includegraphics[width=\columnwidth]{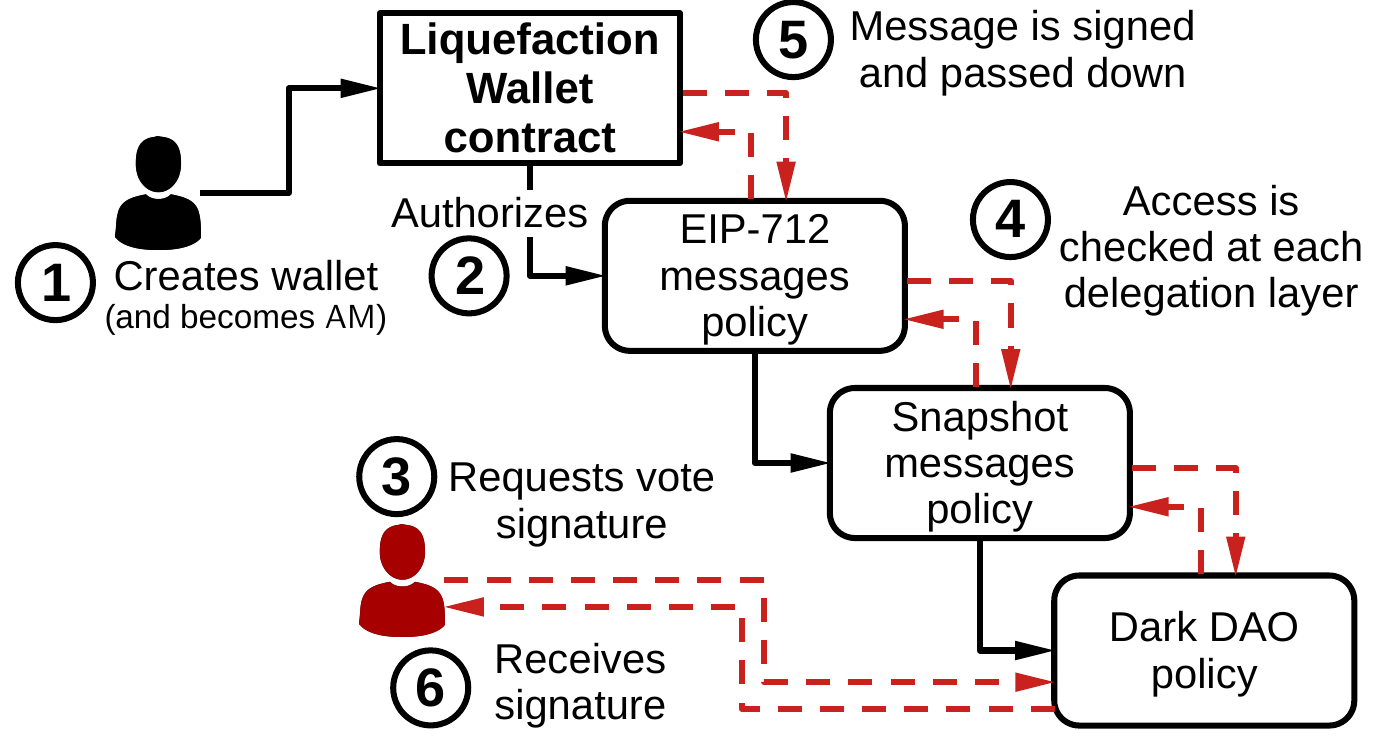}
	\caption{Liquefaction Wallet contract call flow for a Dark DAO policy selling access to vote messages (further described in~\Cref{subsec:governance}). Arrows denote smart contract calls: solid arrows require a transaction, while dotted arrows are performed off-chain. The policies shown are a single path in the delegation tree.}
    \label{fig:liq-wallet-contract-flow}
\end{figure}

\subsection{Wallet Properties}
In this subsection, we discuss a few important properties of \sysname wallets.

\mypara{Policy integrity.}
Due to the architecture of the Ethereum Virtual Machine, policies cannot arbitrarily modify the access given to them. That's because smart contracts cannot directly access or modify the storage of other contracts. The TEE blockchain authenticates all calls to contracts---even those in off-chain simulations---ensuring correct authentication of entities calling a \sysname sub-policy.

\mypara{Transaction types.} A subtle issue in the enforcement of asset-time segmentation is that on-chain smart contracts are arbitrarily expressive. Consequently, there is no way for a \sysname wallet to reason about or rule out policy conflicts for arbitrary on-chain transactions. Instead, \sysname constrains wallets to signing any of three common message types used in the ecosystem of EVM blockchains: Ethereum transaction messages~\cite{EIP1559}, ERC-191 standard messages~\cite{ERC191}, and EIP-712 typed data~\cite{EIP712}. The result is constrained transaction semantics for \sysname wallets, ensuring that \sysname can efficiently identify policy conflicts. (Extensions to other transaction types are possible, but require care to ensure against policy conflicts.)

\mypara{Shared resources.} For conceptual simplicity, we model control of an individual asset as residing with a single player $\player$. As sub-policies are implemented programmatically, however, they can in fact realize more flexible forms of access control. For instance, it is possible to apportion shared access to a single asset in \sysname, e.g., allow simultaneous access by players $\player_1$ and $\player_2$.

\mypara{Address privacy.} Suppose a player $Q$ gains access to a resource in an encumbered wallet---e.g., $Q$ is a briber buying votes in a Dark DAO. $Q$ then can obtain signed transactions from the sub-policy selling the votes, namely transactions casting ballots. These transactions must reveal the address $a$ controlled by the wallet (i.e., corresponding to encumbered key $\sk$). It is possible, however, to conceal $a$ from $Q$ by using a TEE intermediary which publishes signatures / transactions directly to the target blockchain---with careful attention to timing side channels and the size of anonymity set containing $a$.

\subsection{Handling Sub-Balances and Fees}
\label{subsec:transaction-encumbrance-policy}

Asset-time segmentation in \sysname treats fungible assets in a special, flexible way, as discussed in~\Cref{subsec:practical-instantiations}. An address's balance in a particular token or currency, e.g., ether, can be apportioned to different players. The balance is divided into sub-balances (analogous to virtual sub-accounts), each of which is treated as a distinct asset.  
This approach permits concurrent access by multiple applications to a particular asset class in a given address, e.g., to ether in a single encumbered Ethereum address.

We realize sub-balances for ether in encumbered Ethereum accounts through a policy contract in control of all Ethereum transaction messages. Managing per-account ownership of non-ether resources on Ethereum, such as ERC-20 tokens and NFTs, usually involves sending transactions to particular smart contract addresses (e.g., a token's address); these also spend ether. Therefore, in addtion to ether sub-balances, we model transaction destination addresses as separate assets. $\acma$ can create sub-policies and grant them exclusive access to send transactions, spending ether, to one or more particular destination addresses.

\mypara{Goals.} Our policy has to ensure that (1) sub-policies cannot deny service to one another, (2) sub-policies cannot spend funds intended for another sub-policy, (3) additional TEE blockchain transaction costs for signing transactions are minimized, and (4) sub-policy access expires as specified by $\acma$ during sub-policy creation.

\mypara{Pre-signing.} The policy must prevent sub-policies from pre-signing transactions which would enable them to spend funds or maintain continued access after their access expires.\footnote{Even if an authorized destination address does not support directly extracting funds, sub-policies could collude with block builders to extract ether by using a high transaction priority fee.} Although Ethereum transactions do not include expiration timestamps, each account is required to include a sequential transaction nonce in every transaction message it sends~\cite{buterin2014ethereum}. We require all signed transactions to use the current account nonce, which prevents sub-policies from sending more than one transaction after their access expires, fulfilling goal (4) to the maximum extent possible. After a transaction is included on Ethereum, any sub-policy may submit a transaction inclusion proof to increment the account nonce recognized by the policy. This process unlocks signatures for the next transaction and prevents denial of service as required by goal (1). The inclusion proof is a Merkle proof that verifies the transaction is embedded in the transaction trie of a canonical Ethereum block. We assume a trusted oracle of Ethereum block hashes (possible via, e.g.,~\cite{xie2022zkBridge}), to function as the root of trust for these proofs.

\begin{figure}
    \centering
    \includegraphics[width=\columnwidth]{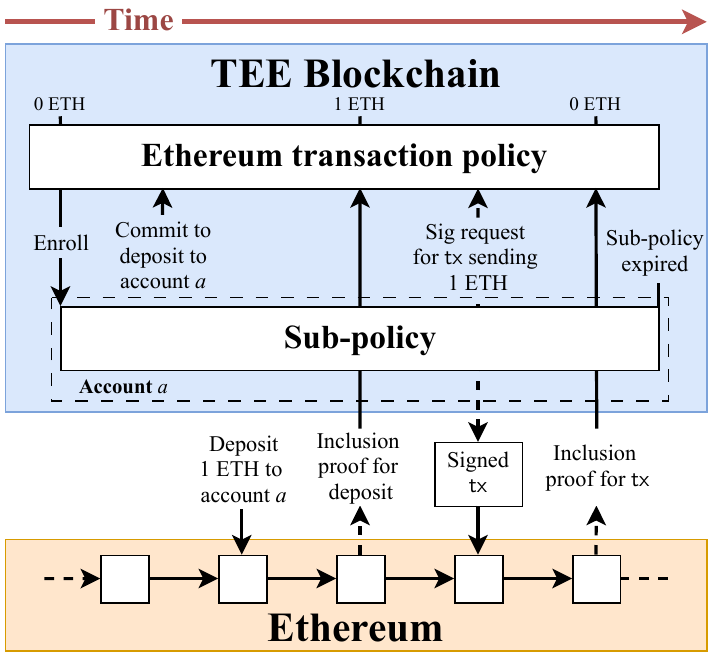}
    \caption{Typical usage of the Ethereum transaction policy: a sub-policy must deposit funds to increase its sub-balance before it can obtain a signed transaction. Solid arrows indicate transactions.}
    \label{fig:eth-transaction-policy-usage}
\end{figure}

\mypara{Accounting.} We maintain an internal mapping from a sub-policy to its ether sub-balance. To deposit funds to its sub-balance, a sub-policy must first create an Ethereum transaction which sends funds to the encumbered account from an outside source and then claim responsibility for it to the policy. The policy enforces that only one sub-policy can claim a particular transaction hash.\footnote{Sub-policies should claim a deposit transaction hash \emph{before} broadcasting the transaction for inclusion on-chain; otherwise, a different sub-policy could learn the transaction hash and claim it first.} After the transaction is included on Ethereum, the sub-policy can submit an inclusion proof to add the deposited funds to its sub-balance. This process is shown in~\Cref{fig:eth-transaction-policy-usage}.

To fulfill goal (2) and prevent sub-policies from spending more than their sub-balance, we enforce a spending limit on each signed transaction. Additionally, we deduct transaction costs from the sender's sub-balance when a transaction inclusion proof is submitted. This deduction occurs atomically with the increment of the recognized account nonce to prevent double-spending.

Aligning with goal (1), we similarly require sub-policies to cover the cost of proving their transactions' on-chain inclusion to the policy. Sub-policies additionally maintain a sub-balance of the TEE blockchain's native token (used for transaction fees on the TEE blockchain) within the policy contract. When a valid transaction inclusion proof is provided to the policy, an estimated cost of this action is deducted from the transaction signer's sub-balance and refunded to the submitter.

When a sub-policy's access ends, leftover funds remain locked in its sub-balance. If $\acma$ grants the sub-policy access again, it can spend funds from its existing sub-balance.

\mypara{Transaction attribution.} Critically, to deduct from the correct sub-balance, the policy must determine which sub-policy signed included transactions. Additionally, we aim to allow sub-policies to sign transactions off-chain (i.e., without requiring changes to the TEE blockchain's state), at no cost, to fulfill goal (3). Typically, we can attribute a transaction $t$ to sub-policy $\Gamma_i$ if $\Gamma_i$ is the sub-policy authorized to sign with $t$'s destination address at the time $t$'s inclusion is proven. However, a special case arises if $\Gamma_i$ waits to broadcast $t$ or prove its on-chain inclusion until \emph{after} its access expires and a second sub-policy $\Gamma_j \ne \Gamma_i$ has been given access by $\acma$ for the same destination address as $t$. This situation can only occur if no other transaction was included before $t$, which would increment the account nonce and invalidate $t$.

To attribute transactions correctly even in this case, we require newly created sub-policies to commit their transaction requests to blockchain storage until the account nonce increases (i.e., until inclusion is proven for a transaction sent from the wallet account). The policy only releases a transaction signature to a new sub-policy after its transaction request is confirmed on the TEE blockchain. This mechanism allows the policy to attribute a transaction $t$ to sub-policy $\Gamma_j$ if either $\Gamma_j$ had committed a request for $t$ or if $\Gamma_j$ was the last sub-policy with unlimited off-chain signing to $t$'s destination address.

\mypara{Performance.} Requiring an inclusion proof to unlock the next transaction introduces a minimum delay between sending transactions which is primarily dependent on the trusted block hash oracle's judgment of block finality. We evaluated the average time between signing an Ethereum transaction through the policy and proving its on-chain inclusion under three block hash oracles with different Ethereum commitment levels recognized by the community~\cite{pavloff23pos}, as shown in~\Cref{fig:eth-tx-policy-e2e-latency}. The strongest assurance against double-spending---block finality---comes at a latency cost of over 17 minutes on average. This latency figure will be substantially reduced by future changes to Ethereum's consensus~\cite{buterin2023ssf}.

We also measured the TEE-blockchain transaction costs associated with using the policy in~\Cref{fig:eth-tx-policy-costs}. All costs are currently orders of magnitude cheaper than the cost of the Ethereum transactions themselves, thereby adding very little additional cost to using the wallet.

\begin{figure}
    \centering
    \small
    \renewcommand{\arraystretch}{1.3} 
    \begin{tabular}{c r r r}
        \hline
        \textbf{Metric} & \textbf{Latest} & \textbf{Justified} & \textbf{Finalized} \\ \hline
        \textbf{Mean (seconds)} & 49.0 & 648.7 & 1,036.9 \\
        \textbf{Standard Deviation} & 7.4 & 125.2 & 113.8 \\ \hline
    \end{tabular}
    \caption{Average end-to-end transaction inclusion and inclusion proof cycle latency across 160 trials, varied by the trusted block hash oracle's confirmation requirements: the oracle confirming the \textit{latest} block accepts a block as soon as it is received, \textit{justified} when it receives attestations from two-thirds of Ethereum's validator set, and \textit{finalized} when it is one epoch behind the most recent finalized block.}
    \label{fig:eth-tx-policy-e2e-latency}
\end{figure}

\begin{figure}
    \centering
    \small
    \renewcommand{\arraystretch}{1.15} 
    \begin{tabular}{l r c}
        \hline
        \textbf{Oasis Transaction} & \textbf{Gas Usage} & \textbf{Cost (USD)} \\ \hline
        Deploy policy & 7,777,890 & \$0.05382 \\
        Add policy to wallet & 169,145 & \$0.00117\\ 
        Add sub-policy & 87,162 & \$0.00060 \\ 
        Deposit commitment & 50,769 & \$0.00035 \\ 
        Prove deposit inclusion & 364,429 & \$0.00252 \\ 
        Transaction commitment & 140,772 & \$0.00097 \\ 
        Prove transaction inclusion & 395,679 & \$0.00273 \\ \hline
    \end{tabular}
    \caption{Costs of using the transaction encumbrance policy, assuming that 1 ROSE = \$0.06919 (the price as of October 29, 2024) and transactions are priced at 100 Gwei, the Sapphire default.}
    \label{fig:eth-tx-policy-costs}
\end{figure}

\subsection{TEE Security Model}
\label{subsec:tee-security-model}
We consider an idealized model of Oasis Sapphire's trusted execution environment~\cite{pass2017sgx}, treating side-channel issues and platform-level deployment mistakes (see, e.g.,~\cite{chen2019sgxpectre,van2020sgaxe,jeanlouis2023sgxonerated}) as out of scope for our exploration in this paper. We also assume the integrity, i.e., correct execution, and liveness of the TEE blockchain. 

Communications to the TEE blockchain can be observed by a network adversary. The system supports secure channels to application instances, however, and we exclude consideration of side channels resulting from, e.g., analysis correlating Oasis Sapphire traffic with on-chain behavior. (Such side channels can be mitigated through injection of noise, e.g., randomized delays.)

TEEs have a history of side-channel vulnerabilities~\cite{10.1145/3456631, MUNOZ2023103180, 10646750} that violate confidentiality and integrity properties. Within the blockchain community, however, TEEs already are being trusted for practical applications that rely on TEE confidentiality~\cite{gogul_baliga_block_building_2023, oasis_protocol}. Also, automated bug-bounty programs for TEE security can be implemented using frameworks like Sting~\cite{kelkar2024sting} or Keystone~\cite{10.1145/3342195.3387532}, encouraging public disclosure and the ethical use of exploits. Thus we believe that despite the risks, key encumbrance is likely to arise in the wild. As we now explain, moreover, some \sysname use cases only require ephemeral exposure of $\sk$ to a TEE. 

\mypara{Ephemeral key exposure.} While there are reasons for concern about the ability of TEEs to protect secret keys in the long term, certain \sysname applications do not require persistent TEE access to $\sk$. Instead, they can realize a model of ephemeral key exposure to the TEE. 

An example is our dusting-attack mitigation (see~\Cref{provenance}). There, $\sk$ can be $(2,2)$-secret shared. The TEE holds one share and the owner $\player$ of $\sk$ can hold the other, inputting it to the TEE for temporary reconstruction of $\sk$ only to initiate outgoing transactions.

\section{Practical Considerations}
\label{sec:practical-considerations}
Our prototype from Section~\ref{sec:implementation} is fully functional, capable of supporting any application of Liquefaction. Yet, real-world deployment of Liquefaction raises additional concerns---both technical and ethical ones---which we discuss in this section.

\subsection{Challenge: Liveness}\label{sec:liveness}
Because \sysname encumbers key material within TEEs, 
no single party has (complete) knowledge of its keys.
The safety of users' financial assets held in a \sysname wallet therefore depends on the liveness / availability of the underlying TEE-based blockchain---in our prototype, the Oasis Sapphire blockchain.

Liveness is a key property of blockchains. In practice, though, even some of the most popular blockchains, such as Solana and Avalanche, have experienced downtime---as recently as February 2024~\cite{solana-halting, avalanche-halting}.

We have consequently designed \sysname to achieve a notion of safety that is \textit{stronger than that of the underlying TEE-based blockchain}.
In this section we describe the design of a \textit{fallback system} (Figure~\ref{fig:backup}) whose objective is to ensure that keys are not lost even in the case of a catastrophic failure or sustained outage of the underlying, or \textit{primary}, blockchain.
In such a case, the fallback protocol transitions to an external, TEE-based fallback system that protects keys using a fallback key-encumbrance policy. This policy extends the lifecycle of \sysname keys, allowing them to be transferred to a new TEE-based blockchain or released to their access managers, removing their encumbrance.

\mypara{Security goals.} 
\sysname's fallback system can be understood as implementing a \textit{fallback trigger} that causes control of \sysname keys to be transitioned from encumbrance policies on the primary blockchain to \textit{fallback encumbrance policies} on the fallback system in the event of a failure in the primary blockchain.
Ensuring correct encumbrance of keys throughout their lifecycle requires liveness and soundness for both the primary and fallback systems.

Additionally, however, it requires the key security property of \emph{sound fallback}: a transition from primary to fallback systems and encumbrance policies should occur if and only if the primary blockchain experiences a failure as specified by the fallback trigger.

Securing sound fallback is the main technical challenge in the design of \sysname's fallback system. 

Ensuring the soundness and liveness of the fallback system itself also requires careful system design. For clarity of exposition and practicality, however, we focus in what follows on a simple fallback trigger and fallback key-encumbrance policy, as follows:

\begin{itemize}
    \item \textit{Fallback trigger:} Transition to the fallback system occurs iff  a liveness  interface in the \sysname application known as a \textit{sentinel wallet} (having $\Gamma = \textsf{1}, \beta = \textsf{0}$) provably fails to 
    sign messages for a time interval of length at least $T$ (e.g., the industry-standard duration of a week~\cite{ArbitrumFAQ}).
    \item \textit{Fallback key-encumbrance policy:} The fallback system releases a key $\sk$ to its access manager, i.e., removes encumbrance.
\end{itemize}

Of course, these are just examples. Other policies are possible.

\begin{figure}
	\centering
	\scalebox{0.8}{
		\input{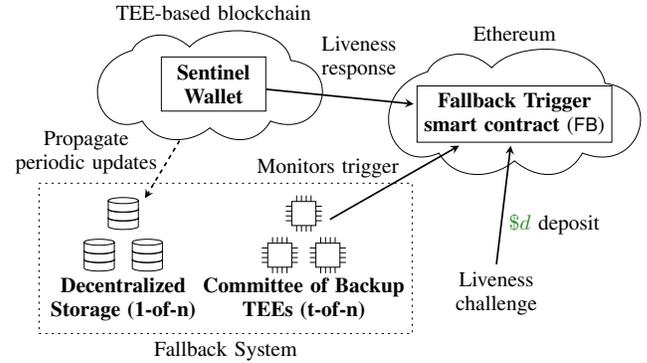}
	}
	\caption{The fallback system is triggered in the rare case that the TEE blockchain goes offline. A smart contract on Ethereum handles challenges to the liveness of the primary system, and a network of backup TEEs take over in the case of a successful challenge.}
    \label{fig:backup}
\end{figure}

\mypara{Design overview.} Briefly, we implement the fallback trigger as a smart contract on a high-reliability blockchain, Ethereum. If at time $\tau$ a liveness challenge $m$ is submitted to the smart contract, and a signature from the sentinel wallet on $m$ is not provided to the smart contract before $T$, the fallback trigger activates. We require a bounty \isgreenmoney{d} to be included with liveness challenges to cover the cost of response transactions to prevent denial-of-service attacks. See Appendix~\ref{sec:appendix-liveness-details} for the full details of our fallback system.

\subsection{Mitigation: Complete Knowledge}\label{sec:complete-knowledge}

\sysname can impact a broad range of applications (Section~\ref{sec:applications}), but applications can prevent mitigate against \sysname by requiring that users provide proofs of \textit{complete knowledge} (CK)~\cite{kelkar2024complete} for their private keys.

A CK proof for a secret key $\sk$ shows that some user has obtained \textit{unencumbered access} to $\sk$ (similar ideas also appeared in~\cite{dziembowski2024individual}). Such proof, once given, applies to a key for the remainder of its lifecycle. Once a key $\sk$ has been disclosed to a user, exclusive possession of the key by an enclave can never be guaranteed and thus can't be proven to a third party.

CK proofs are thus a mitigation to the adversarial applications enabled by \sysname.

We discuss additional details of CK proofs in Appendix~\ref{sec:appendix-complete-knowledge-details}, and other ethical considerations in Appendix~\ref{sec:ethics}.

\subsection{Hiding Liquefaction Wallets}
An important practical consideration for \sysname wallets is the ability to hide information about wallet contents (and sometimes even the existence of \sysname itself) from being publicly provable. For example, if it can be publicly proven that a particular soulbound token is available for rent, this can be used by the token-issuing contract to revoke it, reducing the power of \sysname. In a sense, for correct functioning, the \sysname wallet must be able to prove to potential renters which soulbound tokens are available for rent, without these proofs being transferable. 

The standard way to achieve this is using \textit{designated verifier (DV) proofs}, which are proofs that convince only a designated party $\mathcal{V}$ and no one else. This is accomplished by allowing $\mathcal{V}$ to forge proofs; this prevents $\mathcal{V}$ from convincing another party (including e.g., a public smart contract) by transferring the original proof. 

Remarkably, failure of the SEAO assumption also breaks DV proofs (\cite{kelkar2024complete,mateus2009encumbrance}). Intuitively, if $\mathcal{V}$'s forging key is itself encumbered, then $\mathcal{V}$ regains the power to transfer the proof to others. To get around this, the \sysname wallet can itself weaponize CK to restore the DV property. This will be done by first requesting a CK proof from the verifier $\mathcal{V}$; we call these DV-CK proofs. 

As mentioned before, applications can require CK proofs as a generic solution against key encumbrance. But in the absence of the application requiring CK, \sysname can itself be strengthened by weaponizing CK. 

\section{Blockchain Applications}
\label{sec:applications}

\newcolumntype{a}{>{\columncolor{white}}c}

\newcommand{\fwcell}[2]{%
	\makecell[t]{%
		\begin{minipage}[t]{\getColumnWidth{#1}}
			\raggedright
			#2
		\end{minipage}%
	    \vspace{0.4ex} %
	}
}

\newcommand{\getColumnWidth}[1]{%
	\ifcase#1
	\or
	\or
	\or 4.04cm %
	\or 4.5cm %
	\or 3.09cm %
	\or 2.7cm %
	\or 2.5cm %
	\else 4cm %
	\fi
}

\definecolor{carnelian}{HTML}{B31B1B}

 \begin{figure*}[th!]
    \renewcommand{\arraystretch}{1.5}
    \resizebox{\textwidth}{!}{
    \begin{NiceTabular}{a|clllll}[color-inside]
    \CodeBefore
  \rowcolors{2}{gray!20}{}
\Body
     \toprule
     \rowcolor{gray!50}
     \textbf{Category} & \textbf{Application} & \textbf{Context} & \textbf{\sysname Enables} & \textbf{Broken Assumptions} & \textbf{Constructive Use} & \textbf{Adversarial Use} \\ 
     \midrule
    \multirow{6.5}{*}{Governance} & \multirow{2.4}{*}{\textbf{Voting}} & \fwcell{3}{Decisions in DAOs are made via community votes on proposals~\cite{ethereum-daos}.} & \fwcell{4}{Dark DAOs, which are private DAOs that can operate within public DAOs or systems~\cite{darkdao2018daian}} & \fwcell{5}{Voters control their own tokens.} & \fwcell{6}{Private delegation} &  Bribery attacks \\
    
    & \multirow{2.5}{*}{\textbf{Quadratic Voting}}  & \fwcell{3}{Voting power is distributed quadratically to empower small accounts~\cite{lalley2018quadratic}.} & \fwcell{4}{Using a Dark DAO to square voting power, even in the presence of identity systems} & \fwcell{5}{One person controls each account.} & N/A & \fwcell{7}{Whales increase voting power by account splitting} \\
    
    & \multirow{2.5}{*}{\textbf{Multisigs}} & \fwcell{3}{Multiple signers' approval is required to execute via a multisig~\cite{itakura1983public}.} & \fwcell{4}{Selling access to a key participating in the multisig} & \fwcell{5}{Signers control their own keys.} & \fwcell{6}{Enriched private access structures} & \fwcell{7}{Coordinated theft} \\

    \midrule
    
    \multirow{12}{*}{\makecell[t]{Reputation}} & \multirow{2.25}{*}{\makecell[t]{\textbf{Soulbound} \\ \textbf{Tokens}}} & \fwcell{3}{Soulbound tokens are nontransferable NFTs~\cite{ohlhaver2022decentralized}.} & \fwcell{4}{Renting soulbound token proofs of ownership to third parties} & \fwcell{5}{The account owner has exclusive access to an NFT.} & \fwcell{6}{Facilitating broader group membership}  & \fwcell{7}{Identity misrepresentation} \\

    & \multirow{2.5}{*}{\makecell[t]{\textbf{Transaction} \\ \textbf{History}}}  & \fwcell{3}{Transaction history is a complete and public per-account record~\cite{accounthistory}.} & \fwcell{4}{Purchasing accounts to make deposits in an exchange with strict risk requirements} & \fwcell{5}{The past account user is the current account user.} & \fwcell{6}{Broadening access to credit and service} & \fwcell{7}{Identity misrepresentation} \\

    & \multirow{2.5}{*}{\textbf{Airdrop Rights}} & \fwcell{3}{The right to receive an airdrop can be based on account history~\cite{coinbase_airdrop_2024}.} & \fwcell{4}{Trading encumbered access to accounts that may be eligible for future airdrops} & \fwcell{5}{The past account user is the future account user.} & \fwcell{6}{Unlocking liquidi- ty now by selling future rights}  & \fwcell{7}{Subverting the desired recipients of the airdropper} \\

    & \multirow{2.5}{*}{\textbf{Loyalty Points}} & \fwcell{3}{Loyalty points can be given based on account history~\cite{agrawal2018loyalty}.} & \fwcell{4}{Splitting the cost of a single account which accumulates rewards for a group} & \fwcell{5}{Accounts cannot be shared or transferred.} & \fwcell{6}{Sharing access to points with newer users} & \fwcell{7}{Subverting the desired recipients of the rewards} \\

    & \multirow{2.5}{*}{\textbf{Wash Trading}} & \fwcell{3}{Trading volume for an NFT or token can be artificially increased~\cite{coindesk2022washtrading}.} & \fwcell{4}{Less traceable wash trading} & \fwcell{5}{Trading accounts that appear unconnected are unconnected.} & N/A & \fwcell{7}{Artificial volume which appears authentic}  \\

    \midrule

    \multirow{13}{*}{Privacy} & \multirow{3}{*}{\makecell[t]{\textbf{Trading} \\ \textbf{Locked Tokens}}} & \fwcell{3}{Smart contracts which issue assets sometimes dictate how the assets must be sold or distributed~\cite{lockedtokens}.} & \fwcell{4}{Trading locked tokens and bypassing transfer fees or transfer limits} & \fwcell{5}{Account owners cannot share portions of their accounts.} & \fwcell{6}{Greater token liquidity and more trustworthy derivatives} & \fwcell{7}{Early ``dumping'' of vested tokens, reduced token fee revenue} \\
    
    & \multirow{2.5}{*}{\makecell[t]{\textbf{Private Asset} \\ \textbf{Trading}}} & \fwcell{3}{Standard asset transfers require public, on-chain transactions~\cite{etherscan-understanding-transactions}.} & \fwcell{4}{Trades across multiple user wallets without on-chain transactions} & \fwcell{5}{Account owners control all assets in their accounts.} & \fwcell{6}{Avoiding paying gas for on-chain transactions} & \fwcell{7}{Misrepresentation of activity} \\

    & \multirow{2.3}{*}{\makecell[t]{\textbf{Private DAO} \\ \textbf{Treasuries}}} & \fwcell{3}{DAOs' public treasuries may undermine applications of fundraising DAOs~\cite{enwiki:1225513611}.} & \fwcell{4}{Raising and storing funds in a decentralized, invisible manner} & \fwcell{5}{DAO blockchain  assets are publicly measurable.} & \fwcell{6}{Crowdfunding auction bids} & \fwcell{7}{Attack financing} \\

    & \multirow{2.4}{*}{\makecell[t]{\textbf{Secret Contract} \\ \textbf{Payments}}} & \fwcell{3}{Smart contracts can be made to pay out bounties~\cite{juels2016ring}.} & \fwcell{4}{Paying bounties by revealing the secret keys of encumbered wallets} & \fwcell{5}{Smart contract asset \\ transfers are public.} & \fwcell{6}{Compensating \\ whistleblowers} & \fwcell{7}{Rogue smart \\ contracts} \\

    & \multirow{2.5}{*}{\makecell[t]{\textbf{Hidden Privacy} \\ \textbf{Pools}}} & \fwcell{3}{Privacy Pools is a privacy protocol with built-in accountability~\cite{buterin2024privacypools}.} & \fwcell{4}{A Dark DAO which enforces inclusion of its members in association-set proofs} & \fwcell{5}{Privacy Pools depositors can create any proof.} & \fwcell{6}{Stronger in-group privacy} & \fwcell{7}{Weakened accountability} \\

    \midrule

    \multirow{2.75}{*}{\makecell[t]{Ticketing}} & \multirow{3}{*}{\makecell[t]{\textbf{Token-Gated} \\ \textbf{Ticketing}}} & \fwcell{3}{Access to an event (or metaverse character) can be based on ownership of an asset or NFT~\cite{nftticketing}.} & \fwcell{4}{Transferring event access to someone who does not own the asset} & \fwcell{5}{The account owner has exclusive access to an NFT or asset.} & \fwcell{6}{Unlocking liquidity by renting a ticket- bearing asset} & \fwcell{7}{Undermining event exclusivity or intended audience}  \\

    \midrule

    \multirow{4.5}{*}{Provenance} & \multirow{2.5}{*}{\textbf{Faking Theft}} & \fwcell{3}{When theft happens, people look on chain for evidence~\cite{zhou2023sok}.} & \fwcell{4}{Funds to be sent to a ``thief's'' encumbered account, secretly accessible to the ``victim''}  & \fwcell{5}{Asset transfers establish loss of ownership.} & \fwcell{6}{N/A} & \fwcell{7}{Insurance fraud} \\

    & \multirow{2.5}{*}{\makecell[t]{\textbf{Dusting Attack} \\ \textbf{Mitigation}}} & \fwcell{3}{Tokens from an illicit sou- rce may be sent unsolicited to any other account~\cite{coinbase_crypto_dusting_2024}.} & \fwcell{4}{Proving that the target of a dusting attack has not assumed control of illict assets} & \fwcell{5}{All funds sent to a wallet are accessible to its owner.} & \fwcell{6}{Proving lack of ownership of un- solicited deposits} & \fwcell{7}{N/A} \\

    \midrule

    \multirow{1.75}{*}{\makecell[t]{Cross-chain}} & \multirow{1.75}{*}{\makecell[t]{\textbf{Overlay} \\ \textbf{Smart Contracts}}} & \fwcell{3}{Many blockchains do not support smart contracts~\cite{bartoletti2017empirical}.} & \fwcell{4}{Treating encumbered addresses as smart contract interfaces} & \fwcell{5}{Contracts need native chain support.} & \fwcell{6}{Cross-chain bridges} & \fwcell{7}{N/A}  \\
\bottomrule
\CodeAfter
    \tikz \fill [carnelian] (2-|2) rectangle ($(3-|2)!0.07!(3-|3)$) ; 
    \tikz \fill [carnelian] (5-|2) rectangle ($(6-|2)!0.07!(6-|3)$) ; 
    \tikz \fill [carnelian] (17-|2) rectangle ($(18-|2)!0.07!(18-|3)$) ;  
\end{NiceTabular}
}
\caption{Table of example applications Liquefaction enables or impacts. Rows marked in \textcolor{carnelian}{red} indicate that we have implemented this application.}
\vspace{-0.3cm}
\label{fig:applications}
 \end{figure*}

\sysname undermines the economic and security models of applications that rely on the SEAO assumption. Conversely, it enables new, beneficial applications. In this section, we enumerate the impacts of \sysname across five blockchain application categories: governance, reputation, privacy, ticketing, and provenance. Each application can be realized on \sysname through sub-policy contracts. Our findings are summarized in~\Cref{fig:applications}. In addition, we provide implementations for the Dark DAO, soulbound token sharing, and dusting attack mitigation applications, which we include in our open-source repository.\footnote{\url{https://github.com/key-encumbrance/liquefaction }} We describe several applications in this section, and refer readers to Appendix~\ref{sec:appendix-other-applications} for the details of the rest of the applications shown in~\Cref{fig:applications}. The applications we present are just a few examples of the impact of Liquefaction and are not an exhaustive list of its reach. We believe Liquefaction has a systemic impact on the blockchain ecosystem.

\subsection{Governance}
\label{subsec:governance}
Decentralized Autonomous Organizations (DAOs) are smart-contract-based communities governed by means of voting. 
DAO members cast ballots on proposals to decide which actions the DAO takes. Votes are often weighted by the number of DAO tokens held by an address. Usually, the two capabilities granted by owning a DAO's token---voting power and the token's market value---are intertwined: apart from protocol-supported vote delegation, transferring an individual's voting power requires transferring ownership of their tokens. \sysname enables the separation of these two capabilities which, for ordinary wallets, are inseparable.

\mypara{Dark DAOs.}
We have written an encumbrance policy that liquefies votes in DAOs, realizing a \emph{Dark DAO}, defined in~\cite{darkdao2018daian} as a ``decentralized cartel that buys on-chain votes opaquely.'' Like an ordinary DAO, a Dark DAO is designed to be trust-minimized: it ensures fair exchange, i.e., Alice receives money from a briber if and only if the briber gains agreed-upon access to  Alice's credential(s). Additionally, a Dark DAO is opaque: participation is private, and to a briber, it ensures policy privacy.

A briber desiring a given election outcome (e.g., a ``yes'' vote) can simply offer payment for votes toward this outcome, where payments are scaled to the weight associated with a given bribee's vote. This fair exchange can be implemented using two contracts: an encumbrance policy which broadly encumbers voting messages, and another corresponding to the specific vote-buying contract.

Our voting message encumbrance policy assumes exclusive control of all message signatures pertaining to a popular off-chain DAO voting platform, Snapshot~\cite{snapshottechnicaloverview}. The policy allows a \textit{voting manager} account (by default, the wallet's access manager) to delegate the capability of signing vote messages to exactly one account per DAO proposal. The voting manager may delegate the capability to a vote-buying contract, forgoing the ability to sign voting messages itself. This mechanism guarantees to the Dark DAO that a user cannot override a vote that it signs on the user's behalf.

The briber offers payment for votes in a vote-buying contract which verifies the user has voting power in the DAO and (privately) delegated signing rights to the Dark DAO. After these checks, the contract pays the user by reserving funds which the user may later claim. The entire operation---private voting delegation, voting power verification, and payment---takes place \textit{atomically}, such that if any operation fails, the entire transaction reverts.

\mypara{Tokenized Dark DAO Lite.}
\label{subsec:dark-dao-lite}
We also designed a second Dark DAO system, which we call a \textit{Dark DAO ``Lite''}, in a way that achieves greater usability and coordination, but weaker confidentiality. The key idea behind the Dark DAO Lite is its use of a DAO-token derivative, the \textit{DD token}, to hide the complexity of participation. DD tokens are derived from ordinary target-DAO tokens through a conversion process, and like regular DAO tokens they can be traded on existing token markets, such as Uniswap.

The Dark DAO Lite separates the capabilities of a DAO token into two assets, as follows:

\begin{itemize}
    \item \textit{Voting rights} corresponding to converted target-DAO tokens. A pool of these voting rights may be purchased by a vote-buyer / briber through an auction mechanism. We refer to the pool of voting rights as \textit{self-auctioning}, since the auction process is automated and requires no intervention by DD token holders.
    \item \textit{DD tokens}, which may be individually owned and correspond to ownership rights in the target DAO plus the right to receive revenue from the auctioning of the pool of fractionalized voting rights.
\end{itemize}

Through the use of the Dark DAO ``Lite,'' participation in a Dark DAO is as simple as purchasing DD tokens. Due to the self-auctioning mechanism, the value of held DD tokens will increase relative to the underlying target-DAO token.

See our \href{https://github.com/key-encumbrance/liquefaction }{code repository} for implementation details.

\subsection{Reputation}
Cryptographic keys are often tied to reputation; this is most apparent when an identity is associated with a particular key. The keys' activities---those messages and transactions signed by them---are then assumed to have been performed by the associated identity. Markets can be created which sell either ownership to assets held by a key-encumbered wallet or access to the private key itself, since the TEE controls the private key.

\mypara{Soulbound tokens.}
Soulbound tokens~\cite{Soulbound:2022} are presumed to be illiquid under the SEAO assumption because they are designed to be non-transferrable between accounts. Advocates posit that soulbound tokens will represent ``commitments, credentials, and affiliations'' of individuals in a decentralized society, becoming the standard interface for measuring reputation. Due to their non-transferablilty, soulbound tokens have been suggested as a powerful primitive for reputation~\cite{Soulbound:2022}. However, if the soulbound token is held by an encumbered account, \sysname enables proofs of its ownership to be rented to third parties.

To demonstrate that soulbound tokens are already at risk of encumbrance-based sales, we collaborated with a company (Flashbots) to encumber a soulbound token. Flashbots had recently released a collection of soulbound tokens to its employees. They issued a soulbound token from their collection to an address of our choosing---an encumbered account created from a \textit{maximally upgradable} wallet contract,\footnote{The wallet contract can be found at \href{https://abi-playground.oasis.io/?network=23294&contractAddress=0xF4D023921c910357e7c4964dA0f7666832B526C4}{0xF4D0...26C4} on the Oasis Sapphire Mainnet, and the (previously) encumbered key at \href{https://etherscan.io/address/0x2D25Ab866c9A7db174f3e29961690Ec4d9aE569B}{0x2D25...569B} on Ethereum.} which generates key material but does not support signing messages ($\Gamma = \textsc{0}, \beta = \textsc{0}$). The only function of this wallet is to transfer its key securely to a new wallet contract via a key exchange, thereby making it universally upgradable to any other wallet, encumbered or not.
We discussed with Flashbots how this wallet could then be upgraded to sell access to the soulbound token.
To prevent this, however, we instead released the key by ``upgrading'' the wallet via encrypted transfer to a real Flashbots employee. We plan to work with Flashbots to perform a ``CK ceremony'' among their employees and thereby prove that none of their soulbound tokens are encumbered any longer.

\subsection{Privacy}
\label{subsec:privacy-app}
Due to the public nature of transparent blockchains, the balance and transaction activity of all accounts are public. Public records are not always desirable, however; Liquefaction can facilitate private transfers of ownership that do not leave an on-chain record.

\mypara{Trading locked tokens.}
Some tokens are distributed and activated with a vesting schedule to ensure the investors and founders are aligned with a platform's goals. \sysname enables holders of locked tokens to liquefy and sell their tokens without initiating an on-chain transfer, while maintaining the appearance of respecting their vesting schedules. Even without token locks, prominent community members with publicly tracked wallets might similarly wish to trade their assets without  investor scrutiny. Though some smart contract wallets can already bypass token locks, they do not offer any confidentiality and can be easily blocked by limiting distribution to EOAs and approved wallet contracts. \sysname makes this countermeasure ineffective.

\subsection{Provenance}
\label{provenance}

\mypara{Dusting attack mitigation.}
Most extant blockchain protocols provide no means for users to refuse asset transfers into their accounts. Thus, accounts may be tainted by the unsolicited receipt of funds originating from an illicit source, such as blockchain addresses on a sanctions list---a \emph{dusting attack}~\cite{ofac-dusting}. Victims of dusting transactions from illicit sources may be excluded from business with centralized cryptocurrency exchanges or subject to asset freezes.

Encumbered wallets can provide a defense against dusting. An encumbered wallet which requires asset deposits to be pre-approved before they can be spent can show that the target of the attack does not have, and never had, access to unapproved funds transferred to the account. Our transaction encumbrance policy from \Cref{subsec:transaction-encumbrance-policy} implements this pre-approval mechanism by requiring deposits to be committed before they can be spent. Additionally, upon wallet initialization the access manager can configure this policy to enable exporting proofs of deposit non-ownership.

\section{Related Work}
\label{sec:related}

\mypara{Key encumbrance.} Early work considered a form of key encumbrance in devices with attestation achieved with physical proximity, e.g., visually, and described several (malicious) applications~\cite{mateus09tamperresistance}. A blog post by Daian et al.~introduced the concept of key encumbrance in TEEs with remote attestation, mainly for Dark DAOs---confidential marketplaces for bribing blockchain voters~\cite{darkdao2018daian}. It lacked a TEE-based implementation, however. Li et al.~used key encumbrance to realize offline commitment to transfers of cryptocurrency~\cite{li2021offline}. Related works have explored application-specific encumbrance of web credentials, rather than keys, for internet services such as e-mail and online payments~\cite{matetic2018delegatee} and social-media and government-issued identities~\cite{puddu2019teevil}. Key encumbrance, as implemented through in-account resource locks, has also been suggested as an architecture for enabling atomic, cross-chain state transitions~\cite{onebalance}.

\mypara{Secure credential delegation.} Daian et al.~put forth the notion of a Dark DAO---a DAO that aims to subvert voting in other DAOs--- in~\cite{darkdao2018daian}. In related work, Matetic et al.~propose use of TEEs as a tool for secure credential delegation---which may be unauthorized~\cite{matetic2018delegatee}, and Puddu et al.~explore malicious uses, including  subversion of e-voting~\cite{puddu2019teevil}. Gunn et al. demonstrate the use of remote attestation to undetectably produce provable records of messages that were designed to be deniable, breaking the user privacy guarantees of these services.~\cite{gunn2019circumventing}

\mypara{Encumbrance defense.} Kelkar et al.~proposed and formalized the notion of complete knowledge (CK), the ability to prove that one does not just have access to a secret, but also has fully unencumbered knowledge of it~\cite{kelkar2024complete}. Similar ideas also appeared independently in~\cite{dziembowski2024individual}.

\mypara{Vote-buying / coercion.} There is a considerable literature on the notion of \textit{coercion-resistance} in end-to-end verifiable
voting~\cite{juels2005coercion,delaune2006coercion,lueks2020voteagain}. Broadly speaking, coercion-resistance means that a voter cannot convince a would-be briber or coercer of how she voted. Influential proposed coercion-resistant voting systems include notably Civitas~\cite{clarkson2008civitas} and, more recently, MACI~\cite{Buterin-maci}. None of these definitions or system designs contemplate the risk of key encumbrance; Dark DAOs effectively break all of them.

\section{Conclusion}
\label{sec:conclusion}

\sysname demonstrates the significant impact TEEs and key encumbrance policies can have on the blockchain ecosystem. \sysname shows that breaking the SEAO assumption is feasible by implementing fine-grained access-control policies for encumbered keys. Importantly, \sysname operates confidentially, supporting transactions without on-chain traces and with minimal off-chain visibility.

Our work highlights the limitations of traditional key ownership models for cryptocurrency and the risk of adversaries privately undermining voting integrity, manipulating loyalty systems, and facilitating stealthy wash trading. The obfuscation of transaction history and delegation of control supported by \sysname can enable misuse of DAOs, soulbound tokens, and reputation-based systems. Conversely, \sysname enables new applications, such as privacy-preserving DAOs, overlay smart contracts, and a mitigation to dusting attacks.

Complete Knowledge (CK) offers a countermeasure to key encumbrance, allowing specific applications to avoid its impacts. As TEEs grow more powerful, developers must consider CK in their designs. In summary, our work calls for a broad reconsideration of asset ownership models and recognition of how technologies like TEEs will affect them.

\subsection*{Acknowledgments}
Thanks to Flashbots for productive discussions and collaboration regarding soulbound tokens, and to Andrew Miller for helping carry out the encumbrance of a soulbound token. Thanks also to Amy Zhao for insightful suggestions and comments on this work. This work was funded by NSF CNS-2112751 and generous support from IC3 industry partners and sponsors. Andr\'{e}s F\'{a}brega was funded in part by a TLDR Research Fellowship.

\ifACM \bibliographystyle{ACM-Reference-Format.bst} \fi
\ifUSENIX \bibliographystyle{plain} \fi
\ifIEEE \bibliographystyle{plain} \fi
\ifLNCS \bibliographystyle{Conferences/LNCS/splncs04.bst} \fi

\bibliography{references} 
\ifSP \appendices
        \section{Other Applications}
\label{sec:appendix-other-applications}
This section describes additional applications for \sysname in the blockchain ecosystem.

\subsection{Governance}

\mypara{Quadratic voting.}
Quadratic voting gives greater per-token voting weight to small account holders, thereby reducing the voting power of whales. To win elections under quadratic voting, it is most effective to have a large number of small account holders vote in favor of a proposal. While whales splitting their tokens across Sybil accounts is a known attack on quadratic voting systems, identity systems alone (ensuring one user, one account to negate Sybils) are insufficient in the presence of key encumbrance.  A single whale could \textit{square} its voting power by controlling voting the same number of tokens from a set of other identity-verified encumbered wallets owned by other people.

\mypara{Multisigs.} \textit{Multisignature wallets}, which require signatures from multiple individuals to perform an action, are often used to manage treasuries. Liquefaction could be used to steal treasury funds using a secret coordination scheme among the members of the multisig. To do so, individual signatories secretly signal to a publicly known encumbrance policy their willingness to participate in a theft, and if the signing threshold is met among the conspirators, the encumbrance policy aggregates their signatures and executes the theft. A more prosocial use of Liquefaction in multisigs is to enrich the access structure: for example, a signatory might delegate approving low-risk transactions to an assistant.

\subsection{Reputation}

\mypara{Transaction history.}
Transaction histories are often used to assign risk ratings to blockchain addresses. This allows cryptocurrency exchanges to discriminate against and deny service to users associated with high-risk transaction activity. Since key encumbrance allows the ownership of private keys to be rented or sold, the owner of a blockchain address judged as high-risk might purchase ownership of assets held in a low-risk, key-encumbered wallet and use the low-risk wallet to deposit to a cryptocurrency exchange with strict risk requirements. Transaction risk is therefore liquefied.

\mypara{Airdrop rights.}
Airdrops are a common means of distributing tokens for an emerging protocol. In an airdrop, users receive tokens to their blockchain addresses according to their use of the protocol, community contributions, and other signs of early support. Using key encumbrance, markets can trade encumbered private keys speculated to be eligible for future airdrops. The benefit of trading private keys directly is that the particular form of the airdrop (e.g., knowledge of the airdrop token address or actions required to claim it) need not be known at the time of sale. This allows for flexible redemption and strong guarantees to the key purchaser. By selling rights to a potential future airdrop, early system users can get liquidity earlier by selling their role as potential future token holders to speculators who specialize in selling those tokens at the best moment.

\mypara{Loyalty points.}
In many online services, rewards, special discounts, or exclusive access are granted to users with a history of patronage. For example, certain decentralized exchanges reward users staking their tokens with trading discounts~\cite{1inch:2021, dydxfees:2023}. Using key encumbrance, a group can coordinate to split the cost of a single key which acquires the ongoing rewards and discounts for the whole group, significantly reducing the amount paid to the service while simultaneously making the shared account appear to be a highly active user. Alternatively, a single user can resell discounted access to the service.

\mypara{Wash trading.}
In order to create interest in a particular token, a person or group might engage in illegal \emph{wash trading} to increase the token's trading volume artificially. Key encumbrance can make wash trading harder to detect, since a wash trader could rent access to any number of key-encumbered accounts, each of which many have vastly different provenance and reputation than the others. Given a catalog of enough wallets, the wash trader would be able to evade detection measured by strongly connected components of transfers (as done by \cite{victor2021wash}).

Similarly, one might disrupt NFT provenance by having a celebrity hold the NFT in his or her encumbered wallet to boost the value of the NFT without transferring ownership of the asset.

\subsection{Privacy}

\mypara{Trading assets held in others' wallets.}
Through Liquefaction, one can construct a privacy system that facilitates trading ownership of assets held across multiple accounts. Liquefaction would allow users to create a derivative token that gives access to the full, aggregate balance of an asset held across a series of participating accounts. The derivative would allow users to operate with the full weight of assets held across accounts, for instance to meet the minimum holding requirement to participate in a service, while retaining the privacy of their individual accounts. 

\mypara{Private DAO treasuries.}
Blockchain transparency can sometimes be problematic for DAOs, as illustrated by ConstitutionDAO's attempt to bid over \$40 million at auction on an original first-run copy of the U.S. Constitution~\cite{Chappell:2021}. The DAO's smart contract balance publicly revealed exactly how much money ConstitutionDAO raised, which in part explains how it was narrowly outbid by a hedge fund manager~\cite{Tan:2022}.

Liquefaction enables DAOs to raise money through decentralized treasuries. Using key encumbrance, donations need not even leave the (encumbered) accounts of the donors until they need to be spent---the Liquefaction wallet simply reassigns the ownership of the donated funds to the DAO, which can spend the funds as needed.

\mypara{Secret contract payments.}
Typically, smart contracts explicitly authorize transfers, making them publicly viewable. Liquefaction can be used to facilitate secret payments. A solicitor can put the bounty in an encumbered wallet whose public address is never revealed. The encumbrance contract can prove that there exists a bounty attached to some certain task. Once a user completes this task, the encumbered wallet will reveal the secret key to these funds to the user.

\mypara{Privacy Pools: Hidden sub-pools.}
\textit{Privacy Pools}~\cite{buterin2024privacypools} is a privacy-enhancing protocol that uses a smart contract to offer transaction privacy, much like a mixer such as Tornado Cash. In contrast to a standard mixer, though, Privacy Pools allows users to dissociate their deposits from undesirable addresses and thus from illicit activity. The idea is that the mixing is user-configurable: A user may choose the anonymity set (called an \textit{association set}) that provides privacy for her transaction.

Briefly, on withdrawing funds, a user generates a zero-knowledge proof that the corresponding deposit belongs to a user-selected association set of previous deposits. Users may generate new proofs with reduced association sets as desired, enabling them to exclude deposits from suspect addresses to avoid tainting their own withdrawn funds. 

This feature, however, presents a privacy risk: a depositor may (rightly or wrongly) be subjected to exclusion from other users' association sets and therefore lose the ability to transact privately.

\sysname offers a way to secretly enforce a Tornado Cash-like non-exclusion guarantee. Users can join a Dark DAO that requires association-set proofs to include the deposits of all members, i.e., protects all members against exclusion. While members of such a Dark DAO would risk tainting their transactions, they might be willing to join in the hope of still benefiting from collective privacy or simply for ideological reasons.

Since this application only requires limiting single-user key use, secret sharing (mentioned in~\Cref{subsec:tee-security-model}) can be used to prevent key theft even if the TEE gets compromised.

\subsection{Token-Gated Ticketing}

Some in-person events (e.g.,~\cite{apefest:2024, veecon}) rely on \textit{token-gated ticketing}: admission to the event depends on an attendee's ownership of an asset, such as an NFT, digital identity, or other proof of membership. Though these events are usually billed as exclusive to token holders, Liquefaction can undermine this perceived exclusivity. If the asset enabling entry is held in an encumbered wallet, its owner can delegate admission to the event to someone who does not own the asset. That delegated access could then be redistributed directly or sold in a secondary market.

\mypara{Metaverse ticketing.}
Several video games and virtual worlds (e.g., \cite{apes-come-home-2024, sandbox}) offer features or access exclusively to owners of some NFTs, whose minimum cost of ownership---often referred to as the floor price---may be high. Through Liquefaction, users can rent or lend temporary access to their virtual players without needing to transfer their assets or share a private key. Renting this access offers an entry point for players who want to participate in the virtual world, if temporarily, without taking on the risk or cost of purchasing the requisite NFT.

\subsection{Provenance}

\mypara{Faking theft.}
Encumbrance can also be used to rent \emph{poor} reputation. A malicious entity known to steal private keys and the funds they hold could create a key-encumbered wallet which allows for collusion, given a small bribe. This wallet would be the main recipient of the entity's operations, receiving stolen funds from victims' accounts. However, anyone wishing to fake the theft of their own funds (from a ``fake victim'' wallet) could bribe the key-encumbered wallet in advance in exchange for the ownership of funds sent from the ``fake victim'' wallet. Outside observers would not be able to tell if the fake victim really did have its funds stolen by the entity's operations, but the complicit ``victim'' would have guaranteed future access to their ``stolen'' funds from the recipient wallet.

\subsection{Interoperability}

\mypara{Overlay Smart Contracts.} Since  policies in \sysname are implemented through smart contracts, Liquefaction enables addresses to function as smart contracts on blockchains which do not support smart contracts natively. After encumbering a blockchain address in a Liquefaction wallet, the required contract logic can be implemented as an encumbrance policy. This capability can be beneficial for interoperability frameworks such as asset bridges. A Liquefaction policy can implement the lock-mint functionality that many non-smart contract blockchains lack, thereby supporting safe bridge constructions such as zkBridge~\cite{xie2022zkBridge}.

    \section{Liveness Fallback Details}
\label{sec:appendix-liveness-details}
In this section, we explain how our fallback trigger is implemented and updates are propagated.

\mypara{Implementing the fallback trigger.} 
Figure~\ref{fig:backup} shows the design of \sysname's fallback system, which uses a smart contract \fbcontract to generate a proof that the fallback trigger has occurred, i.e., an outage of duration at least $T$ has occurred. \fbcontract executes not on the primary blockchain (Oasis), but on a separate, high-reliability blockchain, i.e., one whose liveness we assume even if the primary blockchain fails. For this purpose, we use Ethereum, which has experienced only rare, brief outages over its lifetime affecting block finality~\cite{coindesk2023ethereum}.
\fbcontract incentivizes users to generate timely proofs through a challenge mechanism. 
It is initialized with a (large) bounty \isgreenmoney{b}. This bounty may be a one-time deposit per a \sysname instance, as the fallback trigger should be invoked at most once during the lifetime of an instance. On the primary blockchain, support for \fbcontract exists in the form of a special \textit{sentinel} wallet that digitally signs pings off-chain at no cost.  

In the case of an outage on the primary blockchain, at time $\tau$, any user $U_c$ may send a transaction to \fbcontract, along with a deposit \isgreenmoney{d}, that issues a liveness \textit{challenge}. 
Any user may subsequently ping the sentinel wallet, obtaining a signed, timestamped \textit{response} message $r$. The first user $U_r$ to relay such a message $r$ to \fbcontract before time $\tau + T$ is awarded \isgreenmoney{d}, which $U_c$ loses---the fallback trigger proof has failed. Otherwise, if $\tau + T$ passes without a valid response, $U_c$ receives bounty \isgreenmoney{b} and the transition to the fallback system is initiated. (See below for details.) 

Note that the deposit \isgreenmoney{d} can be relatively small---small enough that forfeiting it in the case of mistaken judgment about an outage isn't punitive. Users already have an incentive to ensure that the fallback trigger functions correctly as a way to secure their funds, but the deposit \isgreenmoney{d} additionally protects against denial-of-service attacks on \fbcontract. It makes the cost of submitting many challenge transactions prohibitively expensive and covers the cost of response transactions.

Given the simple fallback key-encumbrance policy we consider here, our fallback system itself has a straightforward and practical design. Rather than requiring a full-blown blockchain, it relies on two sets of services: A fallback committee of \numtees TEE nodes and a set of \numrepos decentralized storage repositories. 

During normal operation of \sysname, \textit{encrypted} copies of system state (both policies and keys) are maintained in each of the decentralized storage repositories. (We give details below.) Copies are encrypted under a public key $\pk_{\sf FB}$ that is distributed using $t$-out-of-$n$ secret-sharing among nodes in the fallback committee upon system setup. 

When \fbcontract successfully proves the occurrence of a fallback trigger, committee members launch a fallback application in a local TEE. This application runs a light client~\cite{ethereum_light_clients}. Upon input of recent Ethereum blocks, it checks the state of \fbcontract to determine whether a valid proof has been created. If so, the committee fetches \sysname state from the decentralized storage repositories. The latest copy is (threshold) decrypted by the committee and instantiated in their TEEs. The application then executes the fallback key-encumbrance policy. 

In the case of the example fallback key-encumbrance policy we consider---removal of encumbrance---the user who generated a private key $\sk$ can export $\sk$ from \sysname. This simple policy requires no communication among nodes in the fallback committee once the fallback application is launched.

\mypara{Propagating updates to the fallback system.}
\sysname must propagate updates to the fallback system periodically to ensure storage of fresh system state. 
The creation of new wallets or decrease of privilege over the encumbered keys must be (encrypted and) propagated to the decentralized storage repositories before the user's actions are considered finalized.
For efficiency, updates that increase privilege over existing keys can be deferred and propagated in batches as the soundness of encumbrance is still preserved in case of fallback. In case of a liveness failure, however, some updates may be missing from the fallback network.

\mypara{Trust model.}
Our security assumptions for the fallback committee and decentralized storage systems are lightweight. We require the availability of at least $t$ nodes upon invocation of a fallback. (They need not be persistently online, but should be launched in response to the failure of the primary blockchain---even if manually.) We additionally require only that one decentralized storage repository provide correct and available data storage.

    \section{CK Proofs Details}
\label{sec:appendix-complete-knowledge-details}
We now expand on our discussion from Section~\ref{sec:complete-knowledge} and give additional details about proofs of Complete Knowledge.

The most practical protocol for a proof of CK itself involves a TEE application. The application outputs $\sk$ locally to the host owner (e.g., displays it on a screen). It can then generate an attestation---on the corresponding public key $\pk$ (or corresponding address $A$)---that such output has taken place. This constitutes a proof of CK. We illustrate how CK might be applied in practice in~\Cref{ex:airdrop}. 
\begin{example}[Airdrop with CK]
\label{ex:airdrop}
Airdrops are executed through smart contracts that control the airdropped assets.
To prevent encumbrance of airdropped tokens, such a smart contract can include an:

\begin{itemize}
    \item On-chain \textbf{CK verification function} that checks a proof of CK sent by a given address $A$ and a
    \item \textbf{Registry} of addresses that have passed CK verification.
\end{itemize} 

The contract then delivers tokens during an airdrop only to addresses present in the registry and therefore known to be unable to encumber tokens. 
\end{example}

As detailed in~\cite{kelkar2024complete}, proofs of CK created from mobile device TEEs are practical today on Ethereum, if expensive (about 1.5 million gas). The cost may drop considerably with native support for the elliptic curves used in attestation signatures~\cite{EIP7212}. Additionally, CK proofs are relatively inexpensive on layer-2 blockchains. We note that while existing mobile-TEE attestations are software based, hardware support is now emerging~\cite{arm_cca_2024}.

    \section{Ethical Considerations}\label{sec:ethics}

Our Liquefaction system has both prosocial and antisocial uses, which we discuss in Section~\ref{sec:applications}. Liquefaction allows accounts to make stronger commitments about their behavior, participate in applications with enhanced privacy, mitigate dusting attacks, and reduce transaction costs. On the other hand, by overturning the SEAO assumption, Liquefaction breaks existing norms that many blockchain application mechanisms rely on.

In traditional security research, it is customary to initiate responsible disclosure for any new software vulnerabilities which are found. However, the disclosure process for open, permissionless systems differs from reporting to traditional software businesses. The lack of central points of contact and the immutability of smart contracts makes complete, private disclosure impossible -- particularly for systemic issues like the SEAO assumption, which affect wide swaths of the blockchain ecosystem rather than specific organizations.

We have adopted a twofold approach to responsible disclosure. First, we privately communicated our findings with major stakeholders, particularly DAOs (Lido, Optimism, and Arbitrum), which we felt were most at risk of attack. We also communicated with vendors such as Oasis and Flashbots, whose platforms could effectively support the hosting of Liquefaction wallets and policies. Second, we have open-sourced the code of our Liquefaction wallet and encumbrance policies. We feel it is important to have a clear demonstration of the practicality of Liquefaction so that the community can understand its long-term implications, especially since practitioners have previously dismissed related works as purely theoretical~\cite{coindesk2018darkdao}.

We also stress that, as explained in~\Cref{sec:complete-knowledge}, proofs of complete knowledge (CK) mitigate Liquefaction threats to systems that adopt them. We hope that our work motivates, and leads to, the development of more practical CK tools which constitute a proactive mitigation.

    \else
\appendix \fi

\end{document}